\documentclass[prb,twocolumn,aps,showpacs]{revtex4-1}
\usepackage{t1enc}
\usepackage[final]{graphicx}
\usepackage{graphicx}
\usepackage{epsfig}
\usepackage{amsmath}       
\usepackage{amsfonts}                                        
\usepackage[utf8]{inputenc}
\usepackage{tikz}                                                      
\usepackage[normalem]{ulem}
\usepackage{etoolbox}
\makeatletter
\patchcmd{\start@aligned}{\null\,}{\null}{}{}
\DeclareRobustCommand*{\bfseries}{
  \not@math@alphabet\bfseries\mathbf
  \fontseries\bfdefault\selectfont
  \boldmath
}
\makeatother
\begin{document}
\title{Domain-wall free energy in Heisenberg ferromagnets}  
\author{Boris Sangiorgio$^{1}$} 
\email{boris.sangiorgio@mat.ethz.ch}
\affiliation{$^1$Department of Materials, ETH Zurich, CH-8093 Zurich, Switzerland}
\author{Thomas C.~T. Michaels$^{2}$}
\email{tctm3@cam.ac.uk}
\affiliation{$^2$Department of Chemistry, University of Cambridge, Lensfield Road, Cambridge CB2 1EW, United Kingdom}
\author{Danilo Pescia$^3$} 
\author{Alessandro Vindigni$^3$}
\email{vindigni@phys.ethz.ch}
\affiliation{$^3$Laboratory for Solid State Physics, ETH Zurich, CH-8093 Zurich, Switzerland} 
\date{\today}                                           
\begin{abstract}
We consider Gaussian fluctuations about domain walls embedded in one- or two-dimensional spin lattices. 
Analytic expressions for the free energy of one domain wall are obtained. From these, the temperature dependence of experimentally relevant spatial scales 
-- i.e., the correlation length for spin chains and the size of magnetic domains for thin films magnetized out of plane -- are deduced. 
Stability of chiral order inside domain walls against thermal fluctuations is also discussed. 
\end{abstract}

\maketitle
\section{Introduction} 
The physics of magnetic domain walls (DWs) has experienced a sort of renaissance during the last decade. 
This was triggered by the perspective of employing DWs in spintronic devices~\cite{Slonczewski_JMMM_96,Parkin_Science_08,Hayashi_Science_08,Allwood_Science_05,Allenspach_PRL08} 
and by the improvement in spatial resolution with which magnetic textures could be resolved~\cite{Wu_PRL_2013,Fischer_Surf-Sci07}.  
A lot of theoretical work has been done to investigate divers physical properties of DWs in view of the novel applicative and 
experimental scenario~\cite{Tatara_PRL04,Ar_Abanov_PRL_10a,Ar_Abanov_PRL_10b,Loss_PRL_12,Yuan_PRL_12}. 
However, the effect of thermal fluctuations within a DW as a single object -- to our knowledge -- has scarcely been investigated~\cite{Nowak_PRB_09,Martinez_PRB,Martinez_JPCM_12}. 
The basic theoretical formalism to tackle this problem analytically was developed between the 70s and the early 80s. 
Then, the thermodynamics of  (not exactly solvable) one-dimensional (1d) classical-spin models was described through a dilute gas of non-interacting DWs,  
including their interplay with spin waves~\cite{Fogedby84JPCSSP,Leung_82,Nakamura_77,Nakamura_78,Schriffer_75}. 
Some results~\cite{Winter_61,Mikeska_JPC_83} have been recently actualized in the context of magnonic applications~\cite{Yan_11,Yan_12,Hertel-Kirschner_PRL04,Bayer_05}.  
Here we focus on the free energy of a single DW, with particular regard to its dependence on temperature and on the system size. We discuss the implications 
on the physics of molecular spin chains~\cite{Miyasaka_review,Coulon06Springer,Bogani_JMC_08,Billoni_11,Gatteschi_Vindigni_13}, ferromagnetic films and nanowires~\cite{Boulle_Mat_SEng_11}. 
Remarkably, the profile of DWs embedded in all these systems is commonly described by the very same model at zero temperature: we consider Gaussian fluctuations~\cite{HB_Braun_PRB94} about this spin profile.  

For the 1d case, the model is presented in Section~\ref{sec_II}, where all the assumptions and analytic results are checked against numerical calculations on a discrete lattice (see also Appendices). 
The strategy followed to compute the DW free energy numerically is explained in details in Section~\ref{sec_III}.  
In Section~\ref{sec_IV} we extend the analytic part to 2d systems. A central result is that Gaussian fluctuations suffice to explain the floating of magnetic-domain patterns and 
the decrease of their characteristic period of modulation with increasing temperature, both facts being observed experimentally~\cite{Oliver_PRL_06,Niculin_PRL_10,Ale_PRB_08}. 
Within the same approximation, we conjecture the absence of chiral order within DWs interposed between saturated magnetic domains. 
\section{The model\label{sec_II}} 
\begin{figure}[b]
\centering
\begin{center}
\includegraphics*[width=8.5cm,angle=0]{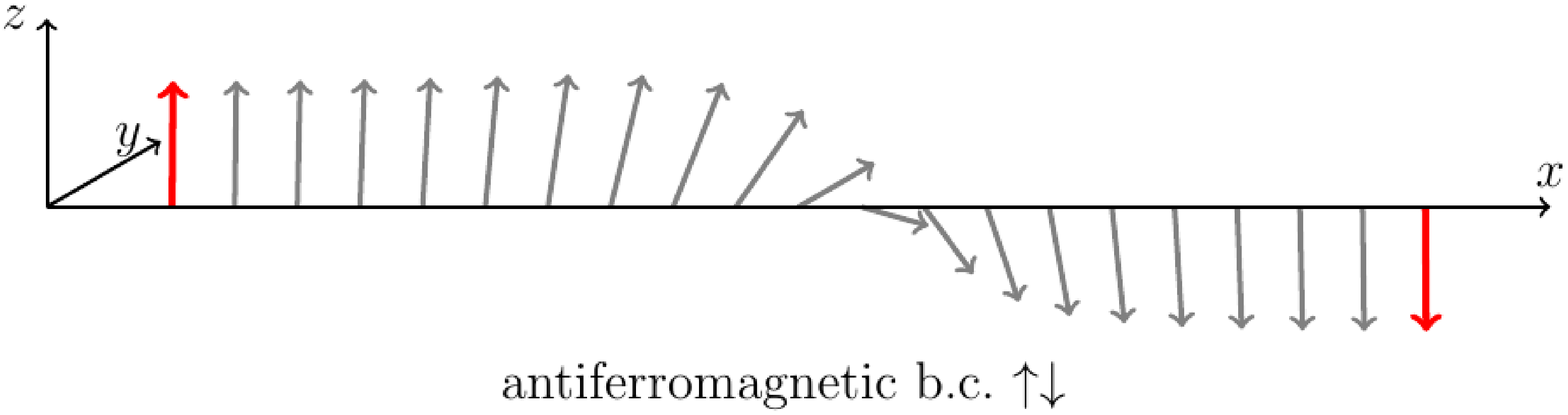}
\vspace{0.5cm}\\
\includegraphics*[width=8.5cm,angle=0]{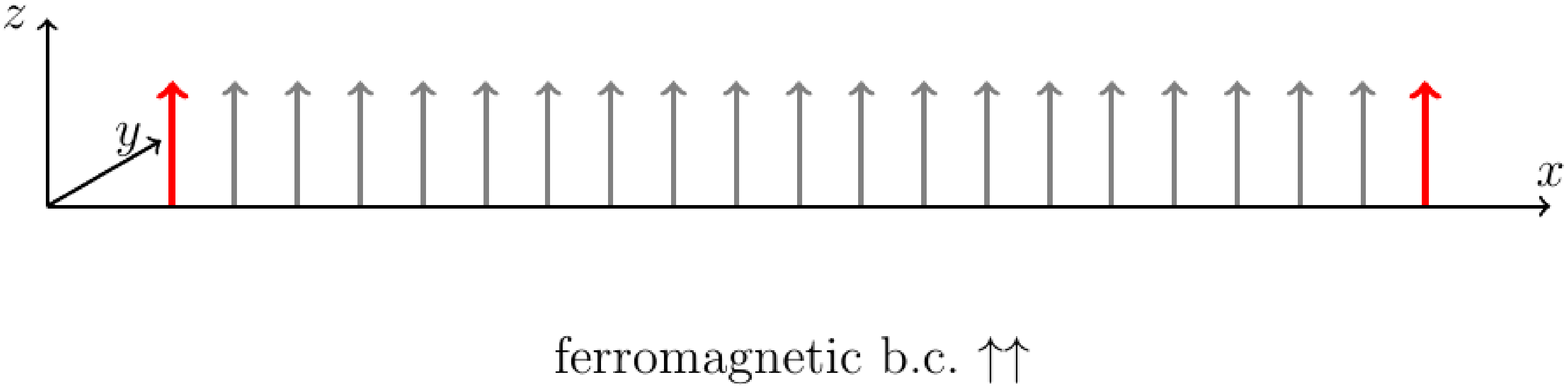}
\end{center}
\vspace{-0.2cm}
\caption{\label{fig1} Sketch of the two boundary conditions considered in the manuscript with the corresponding minimum-energy profile $\vec n(x)$. 
For $\uparrow\uparrow$ b.c. we chose $\vec n(x)=(0,0,1)$, while for $\uparrow\downarrow$ b.c. $\vec n(x)$ is given by Eq.~\eqref{eq18} -- 
with arbitrary $x_0$ and $\varphi_0$ --  and corresponds to the infinite-chain profile.}
\end{figure}
We consider the following classical Heisenberg Hamiltonian:
\begin{equation}\label{eq1}
\mathcal H=- \sum_{i=0}^{L} \left[ J \vec{S}_{i} \cdot \vec{S}_{i+1} + D (S_{i}^z)^2 \right] - D (S_{L+1}^z)^2
\end{equation}
\noindent where $D$ represents the anisotropy energy and $J$ the exchange coupling. Each spin variable $\vec{S}_i$ is a three-component unit vector associated with 
the $i$--th site of the lattice. The two spins at opposite boundaries are forced to lie along the easy anisotropy axis either parallel ($\uparrow\uparrow$) or antiparallel ($\uparrow\downarrow$) to each other (see the sketch in Fig~\ref{fig1}).  
By computing the partition function for these two different boundary conditions (b.c.) the \textit{free-energy} increase associated with the creation of a DW from a uniform ground state can be deduced. 
In this paper we will focus on broad DWs~\cite{Billoni_11}, obtained for $J$ significantly larger than $D$, so that the micromagnetic limit for Hamiltonian~\eqref{eq1} is meaningful:  
\begin{equation}
  \label{eq2}
  \mathcal H=\int\limits_0^{L+1} \mathrm dx\,\left[\frac{J}{2}|\partial_x\vec S|^2 -D\left(S^z(x)\right)^2\right] +\mathrm{const} \,,
\end{equation}
(unitary lattice constant is assumed). 
Ferromagnetic b.c. ($\uparrow\uparrow$) are obtained setting  $\vec S(x=0)=\vec S(x=L+1)=(0,0,1)$, 
while antiferromagnetic b.c. ($\uparrow\downarrow$) correspond to $\vec S(x=0)=(0,0,1)$ and $\vec S(x=L+1)=(0,0,-1)$.  
Following the procedure presented in Refs.~\cite{Polyakov,Politi_EPL_94,Billoni_11}, $\vec S(x)$
is decomposed in two vector fields
\begin{equation}
 \label{eq3}
  \vec S(x)=\vec n(x)\sqrt{1-\vec{\phi}^2}+\vec\phi(x)\,,
\end{equation} 
$\vec\phi(x)$ representing fluctuations and $\vec n(x)$ assumed to vary smoothly in space. 
If $|\vec S(x)|=1$ and $|\vec n(x)|=1$ are required, $\vec n(x)\cdot\vec \phi(x)=0$. Therefore, $\vec\phi(x)$ can be expressed on a \textit{local}, two-dimensional basis orthogonal to $\vec n(x)$
\begin{equation}
\label{eq4}
\vec\phi(x)=\sum_a\phi_a\vec e_a\,,\qquad\mathrm{with}\quad|\vec e_a(x)|=1 . 
\end{equation}
Through the decomposition given in Eq.~\eqref{eq3}, the Hamiltonian~\eqref{eq2} can be expanded for small $\phi_a(x)$ up to quadratic terms, which 
makes it split in two contributions: $\mathcal H= \mathcal H [\vec n] + \mathcal H [ \vec \phi ] +\mathrm{const}$. 
$\mathcal H [\vec n]$ has the same form as Hamiltonian~\eqref{eq2} provided that $\vec S(x)$ is substituted with $\vec n(x)$. 
The fluctuation Hamiltonian reads  
\begin{equation}
  \label{eq5}
\mathcal H [ \vec \phi ] =  \sum_a \int\limits_0^{L+1}\mathrm dx\,\phi_a(x)\hat{H}_a \phi_a(x) \,,
\end{equation}
in which $\hat{H}_a$ acts as a Schr\"odinger--like operator that takes a different form depending on the chosen slow-varying profile $\vec n(x)$. 
After having solved the eigenvalue problem 
\begin{equation} 
 \label{eq6}
  \hat{H}_a\psi(x)=\varepsilon\psi(x)\,,
\end{equation}
each component (labeled by $a$) of the fluctuating field can be expanded on eigenfunctions of $\hat{H}_a$: 
\begin{equation}
  \label{eq7}
  \phi_a(x)=\sum_{\nu}a_{a,\nu} \Psi_{a,\nu}(x)+ \sum_{q} a_{a,q}\Phi_{a,q}(x)\,. 
\end{equation}
In our notation we associate the Greek index $\nu$ with (possible) bound states and $q$ with free states. By free states we mean functions which are delocalized 
throughout the spin chain, also when $L<\infty$. As a consequence of the finite size and of our choice of boundary conditions, such ``free'' states  actually correspond to the wave functions of a free particle in a box when $\vec n(x)$ is assumed uniform (lower sketch in Fig.~\ref{fig1}). In our vocabulary, there is no bound state in this case.    
When $\uparrow\downarrow$ b.c. are assumed, instead, Eq.~\eqref{eq6} admits one bound state per component $a$. As it takes just one value, the label $\nu$ will be dropped henceforth from 
eigenfunctions $\Psi_{a,\nu}(x)$, coefficients $a_{a,\nu}$ and eigenvalues $\varepsilon_{a,\nu}$ 
(quantities associated with free states will still be denoted by the label $q$). 
If $\Psi_{a}(x)$ and $\Phi_{a,q}(x)$ are normalized properly, the expansion~\eqref{eq7} allows rewriting the fluctuation Hamiltonian as 
\begin{equation}
\label{eq8}
\mathcal H[\vec\phi]= \sum_a\left[ \varepsilon_{a} |a_{a} |^2 + \sum_q \varepsilon_{a,q} |a_{a,q} |^2 \right]\,,
\end{equation}
where $\varepsilon_{a}$ and $\varepsilon_{a,q}$ are the bound-state and free-state eigenvalues of Eq.~\eqref{eq6}, respectively.  \\
A partition function that depends parametrically on the slow-varying field $\vec n(x)$ is obtained integrating over fluctuations,  namely  
\begin{equation}
\label{eq9}
\mathcal{Z} [\vec n] =   {\rm e}^{-\beta\mathcal{H} \left[\vec n \right]} \int \mathcal{D}[\vec \phi]\,\,
{\rm e}^{-\beta\mathcal{H} \left[\vec \phi \right]} \,,
\end{equation}
where $\int \mathcal{D}[\phi]$ stands for functional integral and $\beta=1/T$ ($k_B=1$ henceforth). This partition function can be rewritten as product of Gaussian integrals by making use 
of Eqs.~\eqref{eq7} and~\eqref{eq8}:  
\begin{equation}
\label{eq10}
\begin{split}
\mathcal{Z}&[\vec n] = {\rm e}^{-\beta\mathcal{H} \left[\vec n \right]} 
\times\left[\int {\underset{a}{\Pi}}
\mathrm da_{a} {\rm e}^{-\beta\varepsilon_{a} |a_{a} |^2}\right]^{n_{\rm dw}} \\
&\times\int {\underset{a,q}{\Pi}}\mathrm da_{a,q} {\rm e}^{-\beta\varepsilon_{a,q} |a_{a,q} |^2}\,. 
\end{split}
\end{equation}
For antiparallel b.c. ($\uparrow\downarrow$) $n_{\rm dw}=1$, meaning that one DW is present in the spin chain; while $n_{\rm dw}=0$ for parallel b.c. ($\uparrow\uparrow$). 
All the free states have positive energy, which makes their Gaussian integrals convergent.  
After this integration, Eq.~\eqref{eq10} reads  
\begin{equation}
\label{eq11}
\begin{split}
\mathcal{Z}&[\vec n] = {\rm e}^{-\beta\mathcal{H} \left[\vec n \right]} 
\times\left[\int {\underset{a}{\Pi}} \mathrm da_{a} {\rm e}^{-\beta\varepsilon_{a} |a_{a} |^2}\right]^{n_{\rm dw}} \\
&\times\exp\left[-\frac{1}{2} \sum_{a,q}\log\left(\frac{\beta\varepsilon_{a,q}}{\pi}\right) \right]\,. 
\end{split}
\end{equation}
The integration over bound-state amplitudes $\mathrm da_{a}$ needs more care.   
We will see that bound states are associated with vanishing energy so that 
their contribution to the partition function cannot be evaluated through a standard Gaussian integration. The remedy to handle this divergence is presented in details in Ref.~\onlinecite{Leung_82}. 
As the explicit form of the bound-state eigenfunctions $\Psi_{a}(x)$ is required, we prefer to postpone this discussion. 
Further on, we will approximate the summation over free states $\sum_{a,q}$ in Eq.~\eqref{eq11} with an integral. 
This approximation requires the knowledge of the density of states~\cite{Currie_PRB_80}, which also needs to be determined previously by solving the eigenvalue problem in Eq.~\eqref{eq6}.  
From $\mathcal{Z}[\vec n]$ the free energy corresponding the to the b.c. $\uparrow\uparrow$ and $\uparrow\downarrow$ sketched in Fig.~\ref{fig1} can be computed. 
We will focus on the difference between those free energies  
$\Delta F= F_{\uparrow\downarrow} - F_{\uparrow\uparrow}=-T\log\left(Z_{\uparrow\downarrow}/Z_{\uparrow\uparrow}\right)$,  
which we identify with \textit{one} DW free energy. 
Generally, choosing $\uparrow\uparrow$ boundary condition does not warrant the absence of DWs at finite temperatures. 
However, since we work with a finite system, we expect that DWs start forming spontaneously only when $\Delta F \lesssim 0$. 
So we can reasonably think that $\Delta F$ is a good estimate for  \textit{one} DW free energy as long as it is positive.  
\subsection{Schr\"odinger-like eigenvalue problem}  
Using the decomposition in Eq.~\eqref{eq3}, the terms entering Hamiltonian~\eqref{eq2} can be expanded to second order in $\phi_a(x)$, which gives 
\begin{equation}
\begin{split}
  (\partial_x\vec S(x))^2&=(1-\vec{\phi}^2)(\partial_x\vec n(x))^2
  +\sum_a(\partial_x\phi_a)^2 +\sum_ac_a^2\phi_a^2\\
  (S^z(x))^2&=(n^z(x))^2(1-\vec{\phi}^2)+\sum_a\phi_a^2(e_a^z)^2 \,.\label{eq12}
\end{split}
\end{equation}
$c_a$ are the components of $\partial_x\vec n$ on the basis $\vec e_a$ (remember that $|\vec n(x)|=1$ so that $\partial_x\vec n \perp \vec n$).   
The fluctuation Hamiltonian reads
\begin{equation} 
\begin{split}
 &\hat{H}_a=-\frac{J}{2}\partial_x^2 + V_a(x)  \,,\qquad\mathrm{with}\\ 
 &V_a(x) =\frac{J}{2} \left[c_a^2 - (\partial_x\vec n)^2 \right]
 + D \left[(n^z)^2 -(e_a^z)^2\right] \, .\label{eq13} 
\end{split}
\end{equation}
With our choice of boundaries, one has the equivalence 
\begin{equation}
  \label{eq14}
  \int\limits_0^{L+1}\mathrm
  dx\,(\partial_x\phi)^2=-\int\limits_0^{L+1}\mathrm dx\,\phi\,\partial_x^2\phi\,,
\end{equation}
which yields the second derivative in Eq.~\eqref{eq13}.  
The ``potential'' $V_a(x)$ will be specified by the choice of profile $\vec n(x)$.   
The latter will be chosen as the minimal-energy profile consistent with ferromagnetic and antiferromagnetic b.c.. 
These two cases are discussed separately in the following. 
\subsubsection{Uniform profile}  
When ferromagnetic ($\uparrow\uparrow$) b.c. are chosen the ground state is given by a uniform profile. Thus, we set  
$\vec n(x)=(0,0,1)$ or, equivalently, $\vec n=\vec e_z$ and $\vec e_{a}$ given by the two vectors $\vec e_{x,y}$ ($a=x,y$ for this case). 
Accordingly, the potential $V_a(x)$ takes the form
\begin{equation}
  \label{eq15}
  V_a=D +
  \begin{cases}
    0 & x\in (0,L+1)\\
    \infty & \mathrm{else}
  \end{cases}
\end{equation}
so that the eigenvalue problem is formally equivalent to that of a particle in a box whose normalized eigenfunctions are given by
\begin{equation}
  \label{eq16}
  \Phi_{a,q}(x)=\sqrt{\frac{2}{L+1}}\sin(qx)\,, \qquad\mathrm{with} \quad a=x,y
\end{equation}
with $q=l\, \pi/(L+1)$, $l=1,2,3,\ldots$ and eigenvalues 
\begin{equation}
  \label{eq16b}
  \varepsilon_{a,q}=\frac{J}{2}q^2+ D\,.
\end{equation}
%
\subsubsection{Domain-wall profile}  
In the case of antiferromagnetic ($\uparrow\downarrow$)  b.c., $\vec n$ is chosen to be
\begin{equation}
\begin{cases}
  \label{eq17}
  n^x(x)&=\cos\varphi_0 \,{\rm sech}(c(x-x_0))\\ 
  n^y(x)&=\sin\varphi_0 \,{\rm sech}(c(x-x_0))\\ 
  n^z(x)&=-\tanh(c(x-x_0))\,,
\end{cases}
\end{equation}
where $c=\sqrt{2D/J}$  is the inverse DW width. Rigorously, the spin profile in Eq.~\eqref{eq17} minimizes the energy of an infinite chain with  $\uparrow\downarrow$ b.c., 
but here we will use it to build the Schr\"odinger--like equation for a finite chain (see Appendix~\ref{finite_size_DW} for further details). 
As we set no anisotropy in the hard plane ($xy$), nor we consider magnetostatic interaction, DWs parameterized by different angles $\varphi_0$ have the  same energy (e.g., Bloch and N\'eel DWs). 
One of the two vectors $\vec e_{a}$ is the tangent vector $\vec e_{x_0}$
\begin{equation}
\begin{cases}
  \label{eq18}
  e_{x_0}^x(x)&=\cos\varphi_0\tanh(c(x-x_0))\\
  e_{x_0}^y(x)&=\sin\varphi_0\tanh(c(x-x_0))\\
  e_{x_0}^z(x)&={\rm sech}(c(x-x_0))
\end{cases}
\end{equation}
that is proportional to $\partial_x\vec n$:
\begin{equation}
\begin{split}
  \label{eq19}
  \partial_x\vec  n&= -c\,{\rm sech}(c(x-x_0))\,\vec e_{x_0} 
\,,\qquad\mathrm{with}\\
c_{x_0}&=-c\,{\rm sech}\,(c(x-x_0))\,.
\end{split}
\end{equation}
The other one can be found from the vector product of $\vec e_{x_0}$ and $\vec n$: 
$\vec e_{\varphi_0}=\vec e_{x_0} \times\vec n=\left(-\sin{\varphi_0}\,,\,\cos{\varphi_0}\,,\,0\right)$ (note that $c_{\varphi_0}=0$).   
For our choice of profile and b.c., this vector coincides with the DW chirality~\cite{B_Braun_AdvPhys_2012}, that is 
\begin{equation}
\label{eq19b}
\frac{1}{\pi}\int\limits_{0}^{L+1}\vec n\times\partial_x\vec n\, {\rm d} x= - \vec e_{\varphi_0}\,.
\end{equation}
The motivation for labeling the vectors $\vec e_a$ with $a=x_0, \,\varphi_0$ will be clear at the end of this paragraph and -- we believe -- will facilitate reading what follows.  
The components of the fluctuation field $\vec\phi(x)$ on the basis vectors $\vec e_{x_0}$ and $\vec e_{\varphi_0}$ will be labeled accordingly:   
$\phi_{x_0}$ and $\phi_{\varphi_0}$. The spatial dependence of $\vec n(x)$ and $\vec e_{x_0}(x)$ propagates to the potentials entering the Schr\"odinger-like equation 
associated with the two independent components of  $\vec\phi(x)$:
\begin{equation}
  \label{eq20}
  V_{x_0}(x)=V_{\varphi_0}(x)=D\left[2\tanh^{2}(c(x-x_0))-1\right]\,. 
\end{equation}
In the general case in which an intermediate anisotropy axis is present, which breaks the $\varphi_0$ degeneracy in the $xy$ plane, 
the potentials $V_{x_0}(x)$ and $V_{\varphi_0}(x)$ are different. 
When the potentials in Eq.~\eqref{eq20} are inserted into the eigenvalue problem~\eqref{eq6}, one obtains an equation which is a special case of the more general
one: 
\begin{equation}
\label{eq21}
\begin{split}
&\hat{H}_m\psi(x)=\tilde{\varepsilon} \psi(x)\,,  \qquad\mathrm{with}\\
&\hat{H}_m=-\partial_\eta^2 +m(m+1)\tanh^{2}\eta \,, 
\end{split}
\end{equation}
where the change of variable $\eta=c(x-x_0)$ has been performed, the energy $\tilde{\varepsilon}$ is adimensional in units of $D$
and $m \in \mathbb{N}$. 
In terms of the lowering and raising operators 
\begin{equation}
\begin{cases}
&\hat{a}_m  = \partial_\eta + m \tanh \eta \\
&\hat{a}^\dagger_m=-\partial_\eta +  m \tanh \eta \,, 
\label{eq22}
\end{cases}
\end{equation}
one has $\hat{H}_m=\hat{a}^\dagger_m\hat{a}_m +m$. 
In Appendix~\ref{operators} we recall a demonstration~\cite{HB_Braun_PRB94} that if $|\psi\rangle_m$ is an eigenstate of $\hat{H}_m$ then 
$\hat{a}^\dagger_{m+1}|\psi\rangle_m$ is an eigenstate of $\hat{H}_{m+1}$. This property allows us to construct the eigenstates of 
$\hat{H}_{1}$ -- which corresponds to case treated in this paragraph ($\uparrow\downarrow$) -- from the knowledge of the eigenstates
of $\hat{H}_{0}=-\partial_\eta^2$. Operatively, this means that we can obtain the solution to the eigenvalue problem associated with the potentials 
in Eq.~\eqref{eq20} by applying the raising operator $\hat{a}^\dagger_1$ to the eigenfunctions of a free-particle in a box $\psi_0(q,x)=A\cos(qx)+B\sin(qx)$, 
$A$ and $B$ being constants to be specified by b.c.. Details of the calculations are reported in Appendix~\ref{rho}. To proceed in the derivation of the DW free energy 
we only need to know the density of states $\rho(q)$ and the eigenvalues $\varepsilon_{a,q}$. The last few have the same dependence on $q$ as in Eq.~\eqref{eq16b} but the allowed values of 
$q$ are different, which eventually yields a different $\rho(q)$ with respect to the case of ferromagnetic ($\uparrow\uparrow$) b.c..  
In addition to these free states, the Schr\"odinger-like Eq.~\eqref{eq6} (with the potentials given in Eq.~\eqref{eq20}) also admits one bound state~\cite{Fogedby84JPCSSP,HB_Braun_PRB94,Winter_61,Yan_11}:  
\begin{equation}
\label{eq24}
\Psi_{a}(x)=\sqrt{\frac{c}{2}}\,{\rm sech}(c(x-x_0)) 
\end{equation}
with eigenvalue $\varepsilon_{a}=0$ (remember that we drop the index $\nu$ because there is just one bound state per component $a=x_0, \,\varphi_0$). 
When an intermediate-anisotropy term $-D_x (S_{i}^x)^2$ is added to Hamiltonian~\eqref{eq1}, the energy of the bound state $\Psi_{\varphi_0}(x)$ becomes positive. 
This suggests that the vanishing of $\varepsilon_{\varphi_0}$ be related to the degeneracy of the profile 
in Eq.~\eqref{eq17} with respect to the angle $\varphi_0$, when $D_x=0$. In fact, $\partial_{\varphi_0}\vec n= {\rm sech}(c(x-x_0))\,\vec e_{\varphi_0}$, 
meaning that the vector $\vec e_{\varphi_0}$ indicates the direction along which the minimum-energy profile is deformed in response to a variation $\delta \varphi_0$. 
In other words, any rotation on the $xy$ plane of the chiral (vector) degree of freedom -- defined in Eq.~\eqref{eq19b} -- does not affect the DW energy if $D_x=0$; while 
only reflections of the chirality vector, $\varphi_0\rightarrow\varphi_0+\pi$, leave the energy unchanged when $D_x\ne 0$ 
(e.g., left- and right-handed Bolch DWs are degenerate).     
On the contrary, the energy $\varepsilon_{x_0}$ remains zero also when an intermediate-anisotropy axis exists ($D_x\ne0$). This is related to the degeneracy~\cite{HB_Braun_PRB94,Leung_82} with respect to $x_0$, as confirmed by the equivalence $\partial_{x_0}\vec n= c\,{\rm sech}(c(x-x_0))\,\vec e_{x_0}$, obtained from Eq.~\eqref{eq19} with the substitution $x\leftrightarrow x_0$. 
\subsection{Domain-wall free energy} 
In the case in which  $\vec n(x)$ represents a DW profile ($\uparrow\downarrow$ b.c.), 
the considerations exposed above furnish a recipe to replace the integration over the amplitudes $\mathrm da_{x_0}$ and $\mathrm da_{\varphi_0}$ by integrals over the DW center $x_0$ and the angle $\varphi_0$, respectively. 
This makes it possible to evaluate the contribution of bound states to the partition function~\eqref{eq11}.  
From the relations between $\vec e_{x_0,\varphi_0}$, $\partial_{x_0}\vec n$ and $\partial_{\varphi_0}\vec n$ the required Jacobian can be deduced~\cite{Fogedby84JPCSSP,Leung_82}  
\begin{equation}
\begin{split}
\label{eq25}
&|\partial_{x_0}\vec n| \, \mathrm d x_0 = \Psi_{x_0}(x) \mathrm d a_{x_0} \\
&|\partial_{\varphi_0}\vec n|\, \mathrm d \varphi_0 = \Psi_{\varphi_0}(x) \mathrm d a_{\varphi_0} 
\end{split}
\quad\Rightarrow\quad
\begin{split}
&\mathrm d a_{x_0} =  \mathrm d x_0 \sqrt{2c} \\
&\mathrm d a_{\varphi_0} =\mathrm d \varphi_0   \sqrt{2/c} \,.
\end{split}
\end{equation}
The integrals we are interested in reduce to 
\begin{equation}
\label{eq25b}
\int {\underset{a=x_0,\varphi_0 }{\Pi}}  \mathrm da_{a}  = 2  \int\limits_{0}^{L} \mathrm d x_0  \int\limits_{0}^{2\pi} \mathrm d \varphi_0 = 4\pi L\,, 
\end{equation}
where the fact that $\varepsilon_{x_0}=\varepsilon_{\varphi_0}=0$ has been used and the factor 2 comes from the Jacobian deduced in Eq.~\eqref{eq25}. 
Strictly speaking, it would be more appropriate to let the integral over $\mathrm d x_0$ range from $1/(2c)$ to $L+1-1/(2c)$. 
But, since this numerical factors do not affect the results significantly, we prefer to keep analytic expressions as simple as possible. \\
When a uniform $\vec n(x)$ is assumed ($\uparrow\uparrow$ b.c.), the eigenvalue problem in Eq.~\eqref{eq6}   
does not admit bound-state solutions, thus we can formally set the integral in Eq.~\eqref{eq25b} equal to one.  
We approximate the remaining summation on the free states in Eq.~\eqref{eq11} with an integral:  
\begin{equation}
\label{eq26}
\sum_{q}\rightarrow \frac{1}{2}  \int\limits_{-\infty}^{\infty}\,\rho(q) \, \mathrm dq\,.
\end{equation}
The density of states has the form $\rho(q)=\rho_{\uparrow\uparrow}-\gamma(q)$, with 
\begin{equation}
\label{eq27}
\gamma(q) =\frac{2c\tanh(c(L+1)/2)}{\pi(c^2\tanh^2(c(L+1)/2)+q^2)} 
\end{equation}
and $\rho_{\uparrow\uparrow}=(L+1)/\pi$. The density of states for these linear excitations superimposed to a DW profile $\vec n(x)$ is derived 
in Appendix~\ref{rho}. As defined in Eq.~\eqref{eq27}, $\gamma(q)$ corresponds to placing one DW in the middle of the chain ($x_0=(L+1)/2$).  
The general formula would depend parametrically on the coordinate of the DW center (see Eq.~\eqref{eq:362}), which renders the calculation of the partition function unusefully complicated.  
Eventually, the accuracy of all these approximations will be checked  comparing analytic and numerical results (details about the numerical method will be given in Section~\ref{sec_III}).  
Within these assumptions, the partition functions for the two b.c. sketched in Fig.~\ref{fig1} read:     
\begin{equation}
\label{eq28}
\begin{split}
\mathcal{Z}_{\uparrow\uparrow}&= \exp\left[-\frac{1}{2}  \int\limits_{-\infty}^{\infty}\,\log\left(\frac{\beta\varepsilon_{a,q}}{\pi}\right) \rho_{\uparrow\uparrow}\,\mathrm dq \right]\\ 
\mathcal{Z}_{\uparrow\downarrow}&=4\pi L\, \mathrm e^{-\beta \mathcal E_{\rm dw}} \\
&\times\exp\left[-\frac{1}{2}  \int\limits_{-\infty}^{\infty}\,\log\left(\frac{\beta\varepsilon_{a,q}}{\pi}\right)\left(\rho_{\uparrow\uparrow}-\gamma(q)\right)\mathrm dq \right]\,,
\end{split}
\end{equation}
where $\mathcal E_{\rm dw}=2\sqrt{2JD}$ represents the DW energy with respect to the uniform ground state, while $\varepsilon_{a,q}=Jq^2/2 + D$ consistently with Eqs.~\eqref{eq16App} and~\eqref{eq16}. 
Equations~\eqref{eq28} are obtained from Eq.~\eqref{eq11} by setting $n_{\rm dw}=0,1$ for $\uparrow\uparrow$ and $\uparrow\downarrow$ b.c., respectively.  
We are now in the position to give an expression for the DW free energy: 
\begin{equation}
\begin{split}
\label{eq30}
\Delta F&= -T\log\left(Z_{\uparrow\downarrow}/Z_{\uparrow\uparrow}\right)= \mathcal E_{\rm dw} - T\log(4\pi L) \\
&-\frac{T}{2}  \int\limits_{-\infty}^{\infty}\,\log\left(\frac{\beta J}{2\pi} (q^2+c^2) \right) \gamma(q) \mathrm dq \,.
\end{split}
\end{equation}
Using the fact that  
\begin{equation}
\label{eq31_Int}
\int\limits_{-\infty}^{\infty}\,\frac{\log(q^2+c^2)}{q^2+c_L^2} \mathrm dq = \frac{2\pi}{c_L}\log(c+c_L) 
\end{equation}
with $c_L=c\tanh(c(L+1)/2)$, after some algebra, one obtains 
\begin{equation}
\label{eq31}
\Delta F= \mathcal E_{\rm dw} - T\log\left[4\beta D L\left(1+\tanh(c(L+1)/2)\right)^2\right]\,.  
\end{equation}
Note that $\tanh(c(L+1)/2)$ approaches one in the limit $L\gg1/c$. 
\subsection{Polyakov renormalization}  
When ferromagnetic b.c. are considered, the expansion in Eq.~\eqref{eq7} involves only free states $\Phi_{a,q}(x)$, those given in Eq.~\eqref{eq16}. 
Thermal averages of the coefficients $a_{a,q}$ and their products are momenta of the Gaussian integrals entering the partition function $\mathcal{Z} [\vec n]$ and, therefore, fulfill 
the relations 
\begin{equation}
\begin{split}
  \label{eq32}
&\langle  a_{a,q}\rangle = 0\\ 
&\langle  a_{a,q}\, a_{a,q'}\rangle = \frac{T}{2\,\varepsilon_{a,q}} \,\delta_{q,q'} 
\end{split}
\end{equation}
(equipartition theorem). From this it follows that thermal averages of the fluctuation-field components read
\begin{equation}
\begin{split}
\label{eq33}
\langle\phi_{a}^2\rangle_{\uparrow\uparrow} &= \frac{T}{L+1}  \sum_{q} \frac{\sin^2(qx)}{\varepsilon_{a,q}}  \simeq
\frac{T}{2(L+1)} \int\limits_{-\infty}^{\infty}  \frac{\sin^2(qx)}{\varepsilon_{a,q}}  \rho_{\uparrow\uparrow}\,\mathrm dq \\
&=\frac{T}{2\pi J} \int\limits_{-\infty}^{\infty} \frac{1-\cos(2qx)}{q^2 + c^2}\mathrm dq \\
&=\frac{T}{\mathcal E_{\rm dw}} \left[1-\mathrm e^{-2cx} - \mathrm e^{-2c(L+1-x)}\right]\,;
\end{split}
\end{equation}
the first two terms of the last line come directly from integration on the complex plane, while the third one has been added by hand for symmetry with respect to the boundaries 
$x=0$ and $x=L+1$. This contribution is lost when the summation $\sum_{q}$ is approximated with an integral (first line).  
\begin{figure}[b]
\centering
\input{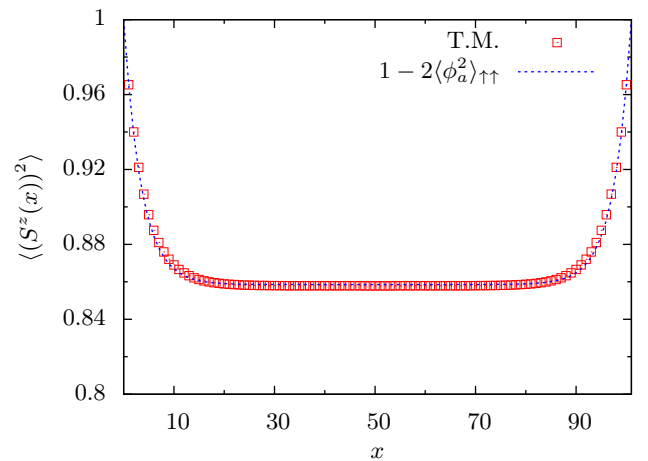} 
\caption{\label{fig2} 
(Color online). $\langle (S^{z}(x))^2\rangle$ obtained for  $\uparrow\uparrow$ b.c.:    
Transfer-matrix calculation (symbols, T.M. in the legend) and analytic result (line) obtained combining Eq.~\eqref{eq33} with Eq.~\eqref{eq34}:  
$\langle (S^{z}(x))^2\rangle= 1-2 \langle\phi_{a}^2\rangle_{\uparrow\uparrow}$. Calculation parameters: 
$D/J = 0.01$, $T/J = 0.02$, $L=100$ spins. }
\end{figure}
Starting from Eq.~\eqref{eq12}, it can be shown~\cite{Billoni_11,Politi_EPL_94} that the following relations hold for thermal averages:  
\begin{equation} 
\label{eq34}
\begin{cases}
\langle(\partial _x \vec{S})^2\rangle &=\left(1-\langle \phi_a^2 \rangle_{\uparrow\uparrow} \right)  \left(\partial _x\vec{n}\right)^2 
+ \sum_{a} \langle\left( \partial_x \phi_a \right)^2 \rangle_{\uparrow\uparrow} \\
\langle\left(S^z(x)\right)^2\rangle & = \left(1-3\langle \phi_a^2\rangle_{\uparrow\uparrow}\right) \left( n^z\right)^2 + \langle \phi_a^2\rangle_{\uparrow\uparrow}\,.
\end{cases}
\end{equation}
In Fig.~\ref{fig2} we compare $\langle\left(S^z(x)\right)^2\rangle$ obtained by inserting the analytic formula for $\langle\phi_{a}^2\rangle_{\uparrow\uparrow}$ given 
in Eq.~\eqref{eq33} with numerical results on a discrete lattice (see Section~\ref{sec_III} for details). The Gaussian approximation agrees very well with numerics at low temperature. 
For an infinite system, the Gaussian approximation would give $\langle\phi_{a}^2\rangle_{\infty}=T/\mathcal E_{\rm dw}$, which is indeed recovered 
away from the boundaries (not shown). 
The more refined, Polyakov approach consists in integrating the free-state degrees of freedom progressively, starting from fluctuations associated with shorter spatial scales. 
Consistently with Eq.~\eqref{eq34}, this leads to renormalization of the spin-Hamiltonian parameters~\cite{Polyakov,Politi_EPL_94} 
\begin{equation} 
\label{eq35}
\begin{cases}
J\rightarrow & J\left(1-\langle \phi_a^2 \rangle_{\mathrm d q} \right) \\
D\rightarrow & D\left(1-3\langle \phi_a^2 \rangle_{\mathrm d q} \right) 
\end{cases}
\end{equation} 
with 
\begin{equation} 
\langle\phi_a^2\rangle_{\mathrm d q}=\frac{T}{2\pi J} \frac{1}{q^2 + c^2}\mathrm dq \,. 
\end{equation} 
The term $c^2$ at the denominator is usually neglected to facilitate the analytic treatment; the error due to this approximation will be compensated by a proper choice of cut-off at which the renormalization procedure stops.  
\begin{figure}[ht]
\centering
\input{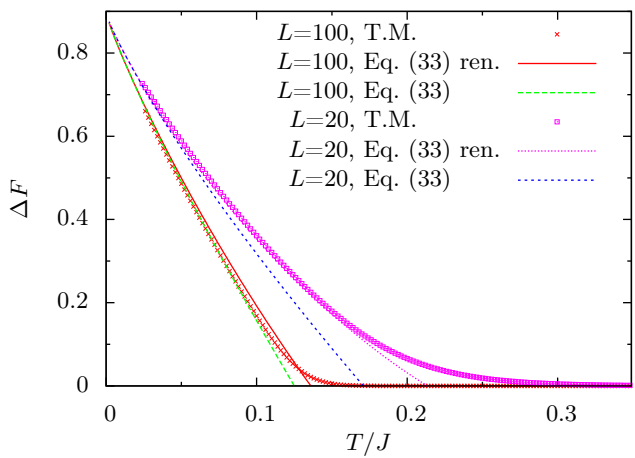}
\caption{\label{fig3} 
(Color online). Domain-wall free energy $\Delta F$ as a function of temperature 
for different system sizes $L$ and $D/J = 0.1$. Symbols (crosses $L=100$ and squares $L=20$) are numerical values obtained by means of transfer-matrix calculations (T.M. in the legend). 
Lines represent the prediction of Eq.~\eqref{eq31}, either using bare constants $D$ and $c$ or renormalized ones $D(\lambda_{\rm c})$ and $c(\lambda_{\rm c})$ (ren. in the legend), as described in the main text. }
\end{figure}
It is convenient to pass to the real space through the substitution $\lambda=1/q$ so that the renormalization flux takes the form 
\begin{equation} 
\label{eq37}
\begin{cases}
&J(\lambda +\mathrm d \lambda)= J(\lambda) \left(1-\frac{T}{2\pi J(\lambda)} \mathrm d \lambda \right) \\
&D(\lambda +\mathrm d \lambda)= D(\lambda) \left(1-3\frac{T}{2\pi J(\lambda)} \mathrm d \lambda \right) \,.
\end{cases}
\end{equation} 
The integration of the first differential equation  is straightforward and gives $J(\lambda)=JZ(\lambda)$, with 
$Z(\lambda)=1-T\lambda/(2\pi J)$ and $J$ being the original, bare exchange constant appearing in Hamiltonian~\eqref{eq1}.  
Taking into account that $\mathrm d Z(\lambda)=-T \mathrm d \lambda/(2\pi J)$, the second equation takes the form 
\begin{equation} 
\label{eq38}
\frac{\mathrm d D(\lambda)}{D(\lambda)} =3\frac{\mathrm d Z(\lambda)}{Z(\lambda)}
\quad\Rightarrow\quad D(\lambda)=D Z^3(\lambda)\,,
\end{equation}
where, again, $D$ is the original, bare anisotropy constant and $D(\lambda)$ the renormalized one. 
The cut-off $\lambda_{\rm c}$ at which the renormalization stops can be chosen in order that the Gaussian result -- consistent with $\langle\phi_{a}^2\rangle_{\infty}=T/\mathcal E_{\rm dw}$ -- 
is recovered at low temperature, namely  $\lambda_{\rm c}=2\pi J/\mathcal E_{\rm dw}=\pi/c$. 
The whole renormalization is meaningful as long as $Z(\lambda_{\rm c})>0$, that is for $T<\mathcal E_{\rm dw}$. 
This is not a problem because the requirement that the DW free energy be positive, $\Delta F \gtrsim 0$, is generally more strict.  \\
In Fig.~\ref{fig3}, the free-energy difference $\Delta F$ computed numerically with the transfer-matrix technique
(see Section~\ref{sec_III}) is compared with theoretical predictions. Two possible system sizes were considered $L=20,\,100$. 
The theoretical curve given in Eq.~\eqref{eq31} and numerical points literally overlap at low temperature. At intermediate temperatures 
it turns out that a better agreement is achieved when $D$ is replaced by $D(\lambda_{\rm c})$ and 
$c$ by $c(\lambda_{\rm c})=c\, Z(\lambda_{\rm c})$ \textit{only} in the argument of the logarithm. 
By increasing $L$, the temperature at which $\Delta F\simeq0$ systematically lowers, as expected from the logarithmic dependence on $L$ in Eq.~\eqref{eq31}. 
For this reason the renormalization effects are less severe and already the Gaussian approximation works very well for $L=100$. However, for $L=20$ the prediction of Eq.~\eqref{eq31} with bare constants
(labeled with ``Eq.~\eqref{eq31}'' the figure legend) deviates more significantly from the exact numerical results. In particular, the transfer-matrix calculation displays a concavity not shown by the theoretical curve obtained with bare values of $c$ and $D$. The concavity is, instead, reproduced at intermediate temperatures by Eq.~\eqref{eq31} if renormalized constants $D(\lambda_{\rm c})$ and $c(\lambda_{\rm c})$ are used (labeled with ``Eq.~\eqref{eq31} ren.'' in the figure legend). 
\subsection{Susceptibility and correlation length }  
In an infinite spin chain pair-spin correlations decay with the distance as  
$\langle S^\alpha(x+r)S^\alpha(x)\rangle=\langle\left(S^\alpha(x)\right)^2\rangle \, \mathrm e^{-|r|/\xi}$, where  $\xi$ is the correlation length and $\alpha=x,y,z$.  
Henceforth, we will focus on their $z$ component and on the magnetic susceptibility along the easy axis, to which they are related through the equation 
\begin{equation} 
\begin{split} 
\label{eq39b}
\chi &= -\frac{1}{L}\frac{\partial^2 F}{\partial B_z^2} = \frac{(g \mu_{\rm B})^2}{T}\langle\left(S^z(x)\right)^2 \rangle\sum_{r=-\infty}^{+\infty}\mathrm e^{-|r|/\xi} \\
&= \frac{(g \mu_{\rm B})^2}{T}\langle\left(S^z(x)\right)^2\rangle \coth\left(\frac{1}{2\xi}\right)
\end{split}
\end{equation}
with $\mu_{\rm B}$ Bohr magneton and $g$ Land\'e factor ($k_B=1$). In the high-temperature limit  $\xi$ vanishes and $\langle\left(S^\alpha(x)\right)^2\rangle$ becomes independent of $\alpha=x,y,z$ so that 
the Curie law is recovered (with the substitution $\langle\left(S^z(x)\right)^2\rangle \rightarrow S(S+1)/3$).  
In the low-temperature limit $\langle\left(S^z(x)\right)^2\rangle \rightarrow S^2$, $\xi \gg 1$ and  $\chi \, T \sim \xi$. The last relation is normally employed to extract information about the spin-Hamiltonian parameters directly from experimental susceptibility data. 
For the Ising chain $\xi\sim \mathrm e^{\beta \mathcal E_{\rm dw}}$ throughout a wide range of temperatures. Therefore, it is generally believed that the thermodynamics of anisotropic spin chains 
may be described with the Ising model just fitting the exchange interaction to the barrier that controls the exponential divergence of $\chi\, T$ ($\Delta_\xi$ in the literature).  
In a previous work~\cite{Billoni_11} we showed that this is not appropriate for the model described by Hamiltonian~\eqref{eq1} when DWs extend over several lattice units. 
This is the regime we address in the present work (the use of Hamiltonian~\eqref{eq2} is legitimate only in this limit, $c\ll1$). 
Existing expansions of $\xi$ often disagree among them and their applicability is limited to some undefined low-temperature region. Here we would like to give an 
analytic expression for $\xi$, based on the interplay between DWs and spin waves, and check it against numerics to establish its actual range of validity.  \\
Figure~\ref{fig3} indicates that Eq.~\eqref{eq31} combined with Polyakov renormalization provides a reasonable estimate for the DW free energy in a finite system. 
Among other applications, this formula can be used to evaluate the behavior of the correlation length $\xi$ as a function of temperature. Let us fix the temperature and define $\bar L$ as the critical length of the spin chain for which $\Delta F=0$. Apart from prefactors, the correlation length is expected to depend on temperature and on the spin-Hamiltonian parameters in the same way as $\bar L$. 
\begin{figure}[ht]
\centering
\input{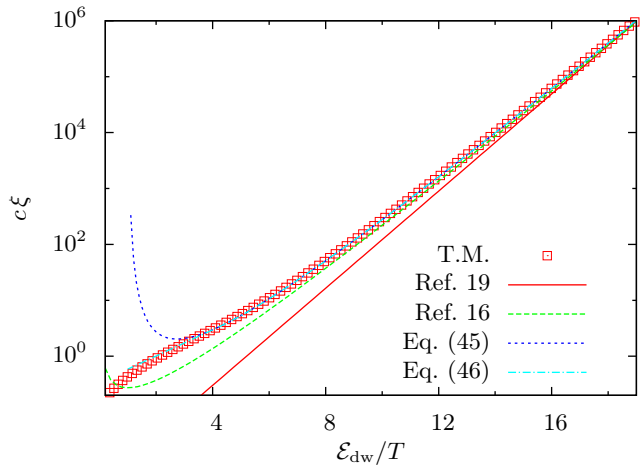}
\caption{\label{fig4} 
(Color online). Correlation length in units of DW width, $c\,\xi$, computed for $D/J=0.1$. 
Symbols are obtained with the transfer-matrix algorithm for an infinite chain (T.M. in the legend).  
Lines represent analytic expressions given in the literature, Eqs.~\eqref{eq40a}--\eqref{eq40b}, and obtained by us. 
The dash-dotted (light-blue) curve (Eq.~\eqref{eq42} in the legend) is obtained by continuing Eq.~\eqref{eq41} with Eq.~\eqref{eq42} for $\mathcal E_{\rm dw}/T<4.262$. The different theoretical curves have been rescaled by constant factors to match the numerical points at low temperatures.}
\end{figure} 
It is convenient to express $\bar L$ in units of DW width. The implicit equation can be solved with the fixed-point method (see Appendix~\ref{xi} for details). 
However, it turns out that the first iteration is accurate within 1.4\% for any $D/J>0.1$. We will, thus, reasonably assume  
\begin{equation} 
\label{eq40}
c\bar L =  \frac{\mathrm e^{\beta \mathcal E_{\rm dw}}}{4 \beta \mathcal E_{\rm dw} Z(\lambda_{\rm c})} 
\end{equation}
as the solution. 
Note that the right-hand side of this equation only depends on $\beta \mathcal E_{\rm dw}$.  
In Ref.~\onlinecite{Billoni_11} it was shown that the correlation length in units of DW width is a \textit{universal} function of the same variable. 
As anticipated, we expect the product $c\,\xi$ to scale like $c\bar L$. In Fig.~\ref{fig4}, numerical results obtained with the transfer-matrix technique are compared with the prediction of 
Eq.~\eqref{eq40} and other two expansions reported in the literature. 
More specifically, in Ref.~\onlinecite{Nakamura_77,Nakamura_78} it was proposed  
\begin{equation} 
\label{eq40a}
\xi \sim \frac{1}{ c }\,\mathrm e^{\beta \mathcal E_{\rm dw}} \,,
\end{equation}
which differs by a temperature-dependent prefactor from the result obtained in Ref.~\onlinecite{Fogedby84JPCSSP}: 
\begin{equation} 
\label{eq40b}
\xi \sim \frac{1}{ c \, \beta \mathcal E_{\rm dw} }\,\mathrm e^{\beta \mathcal E_{\rm dw}}\,. 
\end{equation}
In Fig.~\ref{fig4}, all analytic results have been rescaled by a constant in order to overlap the numerical points at low temperature. 
Equation~\eqref{eq40} needs to be divided by a factor very close to the Euler constant. Our phenomenological treatment is summarized in the low-temperature expansion   
\begin{equation} 
\label{eq41}
\xi \simeq \frac{1}{4 \,\mathrm e  \, c \, \beta \mathcal E_{\rm dw} } \frac{\mathrm e^{\beta \mathcal E_{\rm dw}}}{Z(\lambda_{\rm c})} \,,
\end{equation}
which corresponds to the short-dashed (blue) curve in Fig.~\ref{fig4}. 
Compared with the other two low-temperature expansions, Eq.~\eqref{eq41} reproduces the numerical points up to higher temperatures, but it fails as well 
for $T>\mathcal E_{\rm dw}/4$. More dramatic is the divergence of Eq.~\eqref{eq41} as $T$ approaches $\mathcal E_{\rm dw}$, associated with the vanishing of $Z(\lambda_{\rm c})$. 
We understand this spurious effect as due to the fact that the cut-off $\lambda_{\rm c}$ at which renormalization stops becomes of the order of the system size, $\bar L$.
Tentatively, we tried to tune this artifact by assuming that the contributions arising from spin waves (including Polyakov renormalization) somehow \textit{freeze} when    
$\lambda_{\rm c}$  becomes of the order of the actual number of spins aligned along the same direction:  $\bar L/2 - 3/2c$. This number is evaluated placing, ideally, the DW in the middle of a segment of length $\bar L$; the 
contribution of $1/c$ misaligned  spins lying roughly on half DW is subtracted as well as that of $1/(2c)$ ``blocked'' 
spins at one boundary (consistently with Eq.~\eqref{eq33}).  
One finds that $\lambda_{\rm c}=\bar L/2 - 3/2c$ when $\beta \mathcal E_{\rm dw}=4.262$. 
For higher temperatures, these considerations suggest to replace Eq.~\eqref{eq41} by    
\begin{equation} 
\label{eq42}
\xi=   \frac{A_0}{c} \,\mathrm e^{b \beta \mathcal E_{\rm dw}}\,,
\end{equation}
with constants $A_0= 0.3282$ and $b= 0.5496$ determined by requiring the continuity of $\xi$ and its derivative with respect to $\beta \mathcal E_{\rm dw}$. 
The phenomenological curve obtained using Eq.~\eqref{eq41} for $\beta \mathcal E_{\rm dw}>4.262$ and Eq.~\eqref{eq42} otherwise is plotted with a dash-dotted (light-blue) line in Fig.~\ref{fig4}. 
Though it is not fully justified on rigorous footing, this curve reproduces well the \textit{universal} behavior of $c\,\xi$ versus $\beta \mathcal E_{\rm dw}$ till the correlation length becomes of the order of the 
DW width ($c\,\xi\simeq 1$). At such short scales the notion of correlation length itself becomes somewhat meaningless.  
A compact, analytic expression for the correlation length is provided by Eq.~\eqref{App_xi_4} in Appendix~\ref{xi}. 

Nowadays, the magnetic susceptibility of the model considered here can be obtained numerically essentially with negligible computational effort. 
However, the search for an accurate expansion of $\xi$ has attracted a remarkable interest in the community of Single-Chain Magnets. 
It should be kept in mind that in real systems defects and impurities limit the chain size to $10^2-10^4$ consecutive, interacting spins. 
For typical values of $J$ and $D$ one has $c\simeq0.1$. Therefore, the behavior of the correlation length can be studied experimentally only for $T > \mathcal E_{\rm dw}/6$, for the 
purest spin chains, or  $T > \mathcal E_{\rm dw}/4$, in worst cases. When the latter scenario occurs, Eq.~\eqref{eq42} suggests that the barrier $\Delta_\xi$, defined as 
$\xi\sim\mathrm e^{\beta\Delta_\xi}$, should approach half of the DW energy $\mathcal E_{\rm dw}$. Indeed, this has been observed in Mn(III)-TCNE molecular spin chains~\cite{Ishikawa_12,Balanda_06},  
whose main thermodynamic features should be captured by the model Hamiltonian~\eqref{eq1}. 
In a recently published review article~\cite{Zhang_RSC_2013} it was pointed out that Eq.~\eqref{eq40a} does not succeed in reproducing the measured susceptibility of this family of spin chains 
(though Eq.~\eqref{eq40a} was implicitly attributed to Ref.~\onlinecite{Billoni_11} instead of Ref.~\onlinecite{Nakamura_78}).  

The temperature dependence of $\langle\left(S^z(x)\right)^2\rangle$ can also affect the magnetic susceptibility at intermediate temperatures, $T > \mathcal E_{\rm dw}/4$ (see Eq.~\eqref{eq39b}). According to Eq.~\eqref{eq34} and following ones, also this quantity should depend -- to a large extent -- only on $\beta \mathcal E_{\rm dw}$. 
In passing, we note that within our classical model the correct Curie constant cannot be recovered at high temperature: a fully quantum-mechanical calculation would be needed. Though worth mentioning, 
this question lies beyond the scope of the present work. 
\section{Numerical checks\label{sec_III}} 
In this Section we describe and critically comment the numerical calculations  
that were made in order to check robustness and limitations of our analytic results. 

The derivation of Eq.~\eqref{eq31} strongly relies on the solution to Eq.~\eqref{eq6} and on the correctness of the density of states $\rho(q)$ given in Eq~\eqref{eq27}. 
All these intermediate passages were checked diagonalizing the equivalent discrete-lattice problem, as explained in details in Appendix~\ref{rho}.   

The transfer-matrix algorithm and how finite size constrains the minimal-energy spin profile will be discussed in separate subsections. 
\subsection{Finite-size transfer matrix\label{TransferMatrix}}  
The thermodynamic properties of the model described by Hamiltonian~\eqref{eq1} and -- more generally -- of any  
classical-spin chain with nearest-neighbor interactions can be computed with the transfer-matrix technique. 
This approach has been largely used to derive analytic~\cite{Nakamura_77,Nakamura_78,Scalapino_75} and numerical results~\cite{Blume75PRB,Tannous} for infinite chains. 
Less frequently it has been employed to study finite systems~\cite{Vindigni06APA}, for which more flexible methods, e.g. Metropolis Monte-Carlo, exist.  
However, for our specific problem of computing the free-energy difference between the configurations sketched in Fig.~\ref{fig1}, the transfer-matrix approach
turned out to be extremely efficient. 
Let us separate the bulk interactions from those characterizing the spins at boundaries
\begin{equation}
\label{TM_Hamiltonian_FS} 
\mathcal{H} =-\sum^{L-1}_{i=1}V_{\rm bulk}(\vec{S}_{i},\,
\vec{S}_{i+1})-V_{\rm bc}(\vec{S}_{1})-V_{\rm bc}(\vec{S}_{L})\,. 
\end{equation}
Defining
\begin{equation}
\begin{split}
\label{TM_V_FS} 
&V_{\rm exch}(\vec{S}_{i},\, \vec{S}_{i+1}) = J \vec{S}_{i}\cdot \vec{S}_{i+1} \\
&V_{\rm ss}(\vec{S}_{i}) = D \left(S^z_{i}\right)^2 
\end{split}
\end{equation}
we have
\begin{equation}
\begin{split}
\label{TM_V_FS} 
&V_{\rm bulk}(\vec{S}_{i},\, \vec{S}_{i+1})=V_{\rm exch}(\vec{S}_{i},\, \vec{S}_{i+1}) \\
&\qquad +  \frac{1}{2} \left(V_{\rm ss}(\vec{S}_{i})+V_{\rm ss}(\vec{S}_{i+1})\right) \\
&V_{\rm bc}(\vec{S}_{1})=\frac{1}{2} V_{\rm ss}(\vec{S}_{1}) + J {S}^z_{1} \\
&V_{\rm bc}(\vec{S}_{L})=\frac{1}{2} V_{\rm ss}(\vec{S}_{L}) \pm J {S}^z_{L}\,,
\end{split}
\end{equation}
the last term being plus for $\uparrow\uparrow$ b.c. and minus for $\uparrow\downarrow$ b.c.. 
$V_{\rm exch}$ can possibly describe anisotropic exchange, Dzyaloshinskii-Moriya interaction, etc., and 
$V_{\rm ss}$ any type of single-spin interaction (other anisotropy terms, Zeeman energy, etc.).
With these conventions, the transfer-matrix  kernel~\cite{Wyld} takes the form
\begin{equation}
\label{TM_kernel_FS}
\mathcal{K}_{\rm TM}(\vec{S}_{i},\, \vec{S}_{i+1})=\exp\left[\beta V_{\rm bulk}(\vec{S}_{i},\, \vec{S}_{i+1})\right]  \,.
\end{equation}
The latter defines the following eigenvalue problem: 
\begin{eqnarray}
\label{TM_integral_eq}
\int
\mathcal{K}_{\rm TM} (\vec{S}_{i},\,\vec{S}_{i+1})W_{m}(\vec{S}_{i+1})
d\Omega_{i+1} =\lambda_{m} W_{m}(\vec{S}_{i})
\end{eqnarray}
whose eigenvalues may typically be ordered from the largest to the smallest one:
\[ \lambda_{0}>\lambda_{1}>\lambda_{2}>\ldots \]
The eigenvalue problem~\eqref{TM_integral_eq} can be solved analytically only in few fortunate cases~\cite{Fisher}. 
Generally, it can be converted into a linear-algebra problem and solved numerically by discretizing the unitary sphere~\cite{Stroud,McLaren,Abramowitz}. 
For each temperature, the number of points 
used to sample the solid angle was increased until the desired precision in the free-energy calculation was reached.   
Once that eigenvalues and eigenvectors are known, the partition function is given by
\begin{equation}
\begin{split}
\label{Prtition_TM_FS}
&\mathcal{Z}_{\rm TM}=\sum_{m}\lambda_{m}^{L-1} w_m^1 w_m^L
 \,\,\,\,\,\,\,\,\,\,\,\,\,\,\,\,\,\text{with}
\\
&w_m^1=\int \exp \left[\frac{1}{2}\beta 
V_{\rm bc}(\vec{S}_{1})\right]W_{m}(\vec{S}_{1}) d\Omega_{1}\\
&w_m^L=\int \exp \left[\frac{1}{2}\beta 
V_{\rm bc}(\vec{S}_{L})\right]W_{m}(\vec{S}_{L}) d\Omega_{L}\, .
\end{split}
\end{equation}
The calculation of $Z_{\uparrow\downarrow}$ and $Z_{\uparrow\uparrow}$ just requires to change one sign in the definition of $V_{\rm bc}(\vec{S}_{L})$ given in Eq.~\eqref{TM_V_FS}. 
In this way $\Delta F= -T\log\left(Z_{\uparrow\downarrow}/Z_{\uparrow\uparrow}\right) $ was computed also numerically 
to check the validity of  Eq.~\eqref{eq31}. Results are plotted in Fig.~\ref{fig3}. 
Further details concerning the calculation of $\langle (S^z(x))^2\rangle$ shown in Fig.~\ref{fig2} may be found in Ref.~\onlinecite{Vindigni06APA}, while details about how to 
compute the correlation length for the infinite chain (Fig.~\ref{fig4}) are given in Ref.~\onlinecite{Billoni_11}.
\subsection{General spin profile} 
The whole theoretical treatment developed in the previous and in the next Sections strongly depends on the profile $\vec n(x)$ assumed in Eq.~\eqref{eq17}. 
Strictly speaking, this profile minimizes the DW energy of an infinite chain when the micromagnetic limit is legitimate (i.e., for $c\ll 1$). 
Still working with broad DWs, analytic expressions can be derived for the case of $L<\infty$, which are recalled for the reader's convenience in Appendix~\ref{finite_size_DW}.  
The accuracy of those analytic results was -- in turn -- checked against a discrete-lattice calculation based on non-linear maps~\cite{trallori:94,rettori:95,trallori:96}.  
The most remarkable fact is that when the DW width becomes of the order of the system size ($L$) the dependence of $n^z(x)$ on the spatial coordinate crosses over from hyperbolic tangent to 
cosine (see Fig.~\ref{fig_profile}).  
This imposes a natural range of validity to our model, which depends on the size of the characteristic length involved in a specific calculation. 
For the correlation length, e.g., meaningful analytic results can be derived only as long as  $\xi \gtrsim 1/c$. 
We already commented on this while deriving Eq.~\eqref{eq42}. 
In the next Section we will discuss the temperature dependence of the period of modulation of 2d striped domain patterns. Also in this case, our considerations are expected to hold as long as 
the separation between two successive DWs is much larger than $1/c$. Since the period of modulation is found to decrease with increasing temperature, 
this self-consistency requirement eventually sets a high-temperature threshold above which our theory cannot be applied. 
Remarkably, the mean-field approach prescribes the low-temperature stripe phase to evolve into a single-cosine modulation at the \textit{mean-field} Curie temperature~\cite{Ale_PRB_08,Pighin_PRE_12}.    
However, since the mean-field approximation is known to overestimate the Curie temperature, in realistic systems it may happen that long-range magnetic order is lost before single-cosine modulation 
is attained. Summarizing, if for \textit{some} reason the period of modulation is constrained to be of the order of $1/c$, a single-cosine modulation is expected. Whether this is necessarily the case when 
the true Curie temperature of a magnetic film is approached remains -- to the best of our knowledge -- not clear. 
\section{Domain walls in thin films\label{sec_IV}} 
The same treatment presented up to now can be extended to DWs described by the spin profile in Eq.~\eqref{eq17} embedded in 2d or 3d magnetic lattices. 
Here we focus on 2d systems, realized in thin ferromagnetic films or (flat) nanowires. More concretely, we address systems whose thickness, $\tau$, (defined along the $z$ direction) is smaller than the 
DW width: $\tau<1/c$. 
The generalization of Hamiltonian~\eqref{eq2} to two dimensions reads 
\begin{equation}
\label{eq46}
\mathcal H=\int\limits_0^{L_y+1} \mathrm dy \int\limits_0^{L_x+1}  \left[\frac{J}{2}|\nabla \vec S|^2 -D\left(S^z\right)^2\right] \mathrm dx  \,,
\end{equation}
where $L_x$ and $L_y$ denote the lateral extensions -- in lattice units --  of the system along $x$ and $y$ direction, respectively. 
Exchange and anisotropy constants are related to the atomic counterparts by the thickness: 
$J=\tau J_{\rm atom}$ and $D=\tau D_{\rm atom}$ ($\tau$ being the number of monolayers in the film). 
The vector field defining the local spin direction is a function of two spatial variables $\vec S=\vec S(x,y)$, but it can still  be decomposed as in Eq.~\eqref{eq3}. 
The major difference is that the fluctuation field becomes a function of $x$ and $y$, $\vec\phi(x,y)$, while $\vec n(x)$ is assumed to depend only on $x$. 
Strictly speaking, the assumption of a DW profile being described by Eq.~\eqref{eq17} does not allow considering complex magnetic-domain patterns, such as bubbles or labyrinths.  
Those patterns are not really the focus of the present paper, which rather deals with thermal properties of DWs. 
Being fully determined by $\vec n(x)$, the same basis set $\vec e_a$ defined for the 1d case can be employed to study Gaussian fluctuations about a DW hosted in 
a 2d lattice.    
Generally, the corresponding Schr\"odinger--like operator is obtained by adding a free-particle term, acting along the $y$ direction, to the operators $\hat{H}_a$ defined previously: 
\begin{equation}
  \label{eq47}
\hat{H}_a^{\rm 2d}=  -\frac{J}{2} \partial_y^2+\hat{H}_a \,.
\end{equation}
For the uniform case, $\vec n=\vec e_z$, the eigenfunctions of $\hat{H}_a^{\rm 2d}$ are the same as those of a free particle in a box: 
\begin{equation}
  \label{eq48}
  \Phi_{a,q_x,q_y}(x,y)=\frac{2\,\sin(q_x x)\sin(q_y y)}{\sqrt{(L_x+1)(L_y+1)}}\,,
\end{equation}
with $q_\alpha=l_\alpha \,\pi/(L_\alpha+1)$, $l_\alpha=1,2,3,\ldots$ ($\alpha=x,y$) and eigenvalues 
\begin{equation}
  \label{eq49}
  \varepsilon_{a,q_x,q_y}=\frac{J}{2}\left(q_x^2+q_y^2\right)+ D\,.
\end{equation}
%
When $\vec n(x)$ is chosen as in Eq.~\eqref{eq17}, the free-state eigenfunctions are given by
\begin{equation}
 \label{eq50}
 \Phi_{a,q_x,q_y}(x,y)=\sqrt{\frac{2}{L_y+1}} \sin(q_y y)\Phi_{a,q_x}(x)\,,
\end{equation}
with the same eigenvalues as in Eq.~\eqref{eq49}, $\Phi_{a,q_x}(x)$ being the eigenfunctions defined in Appendix~\ref{rho}. 
The bound-state contribution reads 
\begin{equation}
 \label{eq51}
 \Psi_{a,q_y}(x,y)=\sqrt{\frac{2}{L_y+1}} \sin(q_y y)\Psi_{a}(x)\,,
\end{equation}
with $\Psi_{a}(x)$ defined in Eq.~\eqref{eq24} and eigenvalues $\varepsilon_{a,q_y}=Jq_y^2/2$ (for both components $a=x_0,\varphi_0$).  
By analogy with the 1d case, each component of the fluctuating field $\vec\phi(x,y)$ can be expanded on eigenfunctions of the operator $\hat{H}_a^{\rm 2d}$ defined in Eq.~\eqref{eq47}.  
Finally, the fluctuation Hamiltonian can be written as 
\begin{equation}
\label{eq52}
\mathcal H[\vec\phi]= \sum_a\left[ \sum_{q_y} \varepsilon_{a,q_y} |b_{a,q_y}|^2 + \sum_{q_x,q_y} \varepsilon_{a,q_x,q_y} |b_{a,q_x,q_y}|^2 \right],
\end{equation}
where $b_{a,q_y}$ and $b_{a,q_x,q_y}$ are the projections of a generic $\phi_a(x,y)$ on the basis functions $\Psi_{a,q_y}(x,y)$ and $\Phi_{a,q_x,q_y}(x,y)$, respectively. 
Using the same conventions as in Eq.~\eqref{eq11}, the partition function is now given by
\begin{equation}
\label{eq53}
\begin{split}
\mathcal{Z}&[\vec n] = {\rm e}^{-\beta\mathcal{H} \left[\vec n \right]} 
\times\left[\int {\underset{a,q_y}{\Pi}} \mathrm db_{a,q_y} 
{\rm e}^{-\beta\varepsilon_{a,q_y} |b_{a,q_y} |^2}\right]^{n_{\rm dw}} \\
&\times\exp\left[-\frac{1}{2} \sum_{a,q_x,q_y}\log\left(\frac{\beta \varepsilon_{a,q_x,q_y}}{\pi}\right) \right]
\end{split}
\end{equation}
(the label $\nu$ has been dropped because it takes just one value for each component $a$). 

Let us consider first the contribution of the bound state associated with translation invariance of DW centers, $x_0$. 
We can formally use the substitution defined in Eq.~\eqref{eq25} and replace each amplitude variable by the corresponding  coordinate of DW center 
$b_{x_0,q_y}= x_0(q_y) \sqrt{2c}$ so that 
\begin{equation}
\label{eq54}
\begin{split}
\mathcal{Z}_{x_0}
&=\int {\underset{x_0,q_y}{\Pi}} \mathrm db_{x_0,q_y} 
{\rm e}^{-\beta\varepsilon_{x_0,q_y} |b_{x_0,q_y} |^2} \\
&= (2c)^{L_y/2}\int {\underset{q_y}{\Pi}} \mathrm dx_0(q_y) 
\exp{\left[-\frac{\beta \mathcal E_{\rm dw}}{2} q_y^2 |x_0(q_y) |^2\right]}\,, 
\end{split}
\end{equation}
where the equivalence $\mathcal E_{\rm dw}=2Jc$ has been used. 
The dependence on $q_y$ in $ x_0(q_y)$ is better understood by thinking that in the ground state 
the DW develops as a straight line along $y$. Therefore, $x_0(q_y)$ actually describes the displacement field associated with a corrugation of the DW induced by thermal fluctuations. 
In fact, a 1d Hamiltonian for elastic displacements (with ``stiffness'' $\mathcal E_{\rm dw}$)
\begin{equation} 
\label{eq55}
\mathcal{H}_{\rm el}= \frac{\mathcal E_{\rm dw}}{2}\sum_y\left(x_0(y+1)-x_0(y)\right)^2  
\end{equation}
would yield a partition function equal to Eq.~\eqref{eq54} for $q_y\sim 0$, namely in the continuum-limit formalism. However, 
if the lattice discreteness is reintroduced, undesired divergences are avoided.  From textbooks it is known that Hamiltonian~\eqref{eq55} reads 
\begin{equation} 
\label{eq56}
\mathcal{H}_{\rm el}= 2\mathcal E_{\rm dw} \sum_{q_y}\sin^2\left(\frac{q_y}{2}\right)|x_0(q_y) |^2  
\end{equation}
in the appropriate Fourier space with   $q_y=l_y \,\pi/(L_y+1)$ and $1\le l_y\le L_y$. 
When the Boltzmann weights in  Eq.~\eqref{eq54} are replaced with $\beta\mathcal{H}_{\rm el}$, after Gaussian integrations with respect to $\mathrm dx_0$, one obtains
\begin{equation}
\label{eq57}
\mathcal{Z}_{x_0} = (2c)^{L_y/2}  \exp{\left\{-\frac{1}{2} \sum_{q_y}\log\left[\frac{2\beta \mathcal E_{\rm dw}}{\pi}\sin^2\left(\frac{q_y}{2}\right) \right] \right\} }\,. 
\end{equation}
For $L_y\gg1$, one has 
\begin{equation}
\label{eq58}
\begin{split}
\sum_{q_y}\log\left[\sin\left(\frac{q_y}{2}\right)\right]&\simeq \frac{L_y+1}{\pi} \int\limits^{\pi}_{0}\log\left[\sin\left(\frac{q_y}{2}\right)\right] {\rm d} q_y\\
&=-(L_y+1)\log2
\end{split}
\end{equation}
which leads to 
\begin{equation}
\label{eq59}
 \mathcal{Z}_{x_0} =  \left(\frac{2\pi}{\beta J }\right)^{L_y/2}\,.
\end{equation}
(within the present approximation $L_y+1\simeq L_y$). 

With a similar strategy, the contribution due to the bound state with vanishing energy associated with degeneracy in $\varphi_0$ can be evaluated. 
Still from Eq.~\eqref{eq25} it follows that $b_{\varphi_0,q_y}= \varphi_0(q_y)\sqrt{2/c}$. The corresponding factor in the partition function is transformed accordingly 
\begin{equation}
\label{eq60}
\begin{split}
\mathcal{Z}_{\varphi_0}
&=\int {\underset{\varphi_0,q_y}{\Pi}} \mathrm db_{\varphi_0,q_y} 
{\rm e}^{-\beta\varepsilon_{\varphi_0,q_y} |b_{\varphi_0,q_y} |^2} \\
&= \left(\frac{2}{c}\right)^{L_y/2}\int {\underset{q_y}{\Pi}} \mathrm d\varphi_0(q_y) 
\exp{\left[-\frac{\beta J}{c} q_y^2 |\varphi_0(q_y) |^2\right]}\,.
\end{split}
\end{equation}
Establishing the same analogy as before, with a ``stiffness'' $2J/c$, a contribution equivalent to Eq.~\eqref{eq59} is recovered, i.e.  $\mathcal{Z}_{\varphi_0}= \mathcal{Z}_{x_0}$. 
On a more physical basis, the Boltzmann weight in Eq.~\eqref{eq60} may be thought of as arising from an XY-model Hamiltonian 
\begin{equation} 
\label{eq61}
\begin{split}
\mathcal{H}_{\rm xy}&= -\frac{2J}{c}\sum_y\cos\left[\varphi_0(y+1)-\varphi_0(y)\right] + \frac{2J}{c} L_y \\
&= -\frac{2J}{c}\sum_y\vec e_{\varphi_0}(y+1)\cdot\vec e_{\varphi_0}(y) + \frac{2J}{c} L_y 
\end{split}
\end{equation}
involving the $\varphi_0$ angles of different arrays of spins, labeled by $y$. From the second line it is clear that $\mathcal{H}_{\rm xy}$ describes the effective coupling between chirality vectors of neighboring chains.  
Along this line, Eq.~\eqref{eq60} can equivalently be written as 
\begin{equation}
\label{eq62}
\begin{split}
\mathcal{Z}_{\varphi_0}&\simeq \left(\sqrt{\frac{2}{c}}\, {\rm e}^{-2\beta J/c}\right)^{L_y} 
\int\limits_{-\pi}^{\pi}\mathrm d\varphi_0(1) \dots \mathrm d\varphi_0(L_y) \\
& \exp{\left[\frac{2\beta J}{c}\sum_y \cos\left(\varphi_0(y+1)-\varphi_0(y)\right) \right]}\,. 
\end{split}
\end{equation}
In terms of relative angles $\delta_y=\varphi_0(y+1)-\varphi_0(y)$ (and assuming $L_y+1\simeq L_y$) the above expression takes the form
\begin{equation}
\label{eq63}
\begin{split}
\mathcal{Z}_{\varphi_0}&= \left(\sqrt{\frac{2}{c}}\, {\rm e}^{-2\beta J/c}\right)^{L_y} 
\left[\int\limits_{-\pi}^{\pi} \mathrm d\delta_y {\rm e}^{\frac{2\beta J}{c}\cos\left(\delta_y\right) }\right]^{L_y}\\
& = \left[2\pi \sqrt{\frac{2}{c}}\, {\rm e}^{-2\beta J/c} \,
\mathcal I_0 \left(\frac{2\beta J}{c}\right) \right]^{L_y} 
\end{split}
\end{equation}
where $\mathcal I_0(\kappa)$ is the modified Bessel function of the first kind. For $2\beta J/c\gg 1 $ (low temperatures) 
it is $\mathcal I_0(\kappa) \simeq \mathrm e^{\kappa}/\sqrt{2\pi \kappa}$ and $\mathcal{Z}_{\varphi_0}= \mathcal{Z}_{x_0}$ is recovered again.

As for 1d case, we focus on the ratio between the partition functions obtained for $\uparrow\downarrow$ and $\uparrow\uparrow$ b.c.. 
For one DW embedded in a thin film that is   
\begin{equation}
\label{eq64}
\begin{split}
&\frac{\mathcal{Z}_{\uparrow\downarrow}}{\mathcal{Z}_{\uparrow\uparrow}}=  \mathrm e^{-\beta L_y \mathcal E_{\rm dw}} \mathcal{Z}_{x_0} \mathcal{Z}_{\varphi_0}  \\
&\times\exp\left[\frac{1}{2} \sum_{q_y}\int\limits_{-\infty}^{\infty}\log\left(\frac{\beta J (q_x^2+q_y^2+c^2)}{2\pi}\right) \gamma(q_x)\,{\rm d}q_x \right].
\end{split}
\end{equation}
Note that \textit{only} the summation over the $q_x$ label has been substituted with an integral (following the same conventions as in Eqs.~\eqref{eq26} and~\eqref{eq27}). 
Moreover, the fact that the energies $\varepsilon_{a,q_x,q_y}$ in Eq.~\eqref{eq49} actually do not depend on the label $a$ has been used.   
Exploiting again Eq.~\eqref{eq31_Int}, the argument of the exponential in Eq.~\eqref{eq64} can be approximated as 
\begin{equation}
\begin{split}
\label{eq65}
&\sum_{q_y}\left[\log\left(\frac{\beta J}{2\pi}\right) + 2 \log\left(c_L+\sqrt{q_y^2+c^2}\right) \right] \\
&\simeq L_y\log\left(\frac{\beta J c_L^2}{2\pi}\right)  \\
& + \frac{2 (L_y+1)}{\pi}\int\limits_{0}^{\pi} \log\left[1+\sqrt{\left(\frac{q_y}{c_L}\right)^2 + \left(\frac{c}{c_L}\right)^2}\right] \mathrm dq_y
\end{split}
\end{equation}
where we considered $L_y\gg1$ and $c_L=c\tanh(c(L_x+1)/2)$ now. The integral in the last line of Eq.~\eqref{eq65} can be solved analytically 
\begin{equation} 
\begin{split} 
&\int\limits_{0}^{\pi} \dots \mathrm dq_y 
=\pi \, \left[\log\left(1+\tilde c\,\cosh t\right) \right] + c_L\, t -\pi\\
&-2 c_L \sqrt{\tilde c^2 -1} \,\arctan \left[\sqrt{\frac{\tilde c-1}{\tilde c+1}} \tanh\left(\frac{t}{2}\right) \right]\\
&{\overset{L_x\gg1}{=}}
\pi\, \log\left(1+\sqrt{1 + \left(\frac{\pi}{c}\right)^2}\right) + c \, {\rm arcsinh}\left(\frac{\pi}{c}\right) -\pi
\end{split} 
\end{equation}
with $\sinh t= \pi/c$ and $\tilde c= c/c_L$. The latter approaches one when $L_x\gg1$ is considered.  
To the leading order, the free energy per unit length of a DW embedded in a thin film, whose linear dimensions are much larger than the lattice spacing, reads 
\begin{equation}
\begin{split}
\label{eq67}
\mathcal F_{\rm dw} &= 
-\frac{T}{L_y}\log\left(Z_{\uparrow\downarrow}/Z_{\uparrow\uparrow}\right) \\
&=\mathcal E_{\rm dw}  -2\,T\left( \frac{c}{\pi}\,{\rm arcsinh}\left(\frac{\pi}{c}\right) -1\right)\\
&-2\,T\,\log\left[c\left(1+\sqrt{1 + \left(\frac{\pi}{c}\right)^2}\right)\right]
-\frac{T}{L_y}\log(4\pi L_x)
\end{split}
\end{equation}
in which  $\mathcal{Z}_{\varphi_0}= \mathcal{Z}_{x_0}$ has been assumed for the sake of simplicity.
The last entropic term arises from two contributions neglected in $\mathcal{Z}_{x_0}$ and $\mathcal{Z}_{\varphi_0}$, namely a rigid translation of the DW as a whole and a rigid rotation of the chirality vectors. 
Those contributions have been already computed for the 1d case in Eq.~\eqref{eq25b}, for which $y$ takes just one value. 
Even if this entropic term practically does not affect the free energy for $L_x,L_y\gg1$, it is important in relation to the underlying degeneracies with respect to $x_0$ and $\varphi_0$. 
In fact, these degeneracies eventually determine the i) \textit{floating} of magnetic-domain patterns and ii) the vanishing of critical current for DW motion in nanowires with no intermediate anisotropy ($D_x=0$). 
\subsection{Positional order of domain walls} 
\begin{figure}[hptb]
\centering
\includegraphics[width=.45\textwidth]{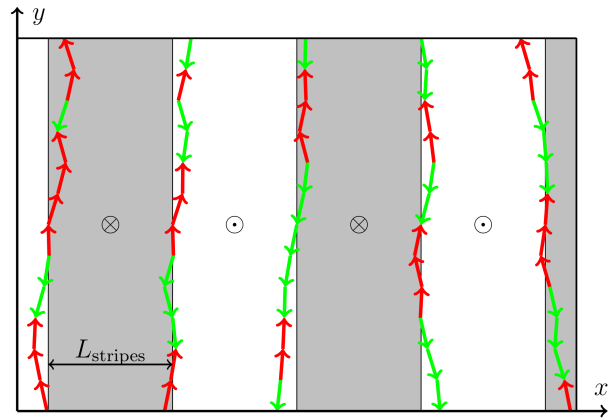}
\caption{Schematic view of stripe pattern in a film with out-of-plane anisotropy. 
Gray and white areas represent domains with opposite magnetization in the ground state. 
Colored (online) arrows give a pictorial representation of DW corrugation and short-range chiral order at finite temperature. 
Note that Bloch DWs have been assumed and disorder in the arrows direction ($\varphi_0(y)$ variable) has been exaggerated to help visualization. }
\label{fig5}
\end{figure}

The fact that both $\mathcal{Z}_{\varphi_0}$ and  $\mathcal{Z}_{x_0}$ arise from elastic-like Hamiltonians has important implications which deserve some further comment. 
As already pointed out, $\mathcal{Z}_{x_0}$ is associated with corrugation of the DW as a function of $y$ (see the sketch in Fig.~\ref{fig5}). In the ideal case considered here, this instability destroys positional order (defined by the profile $\vec n$) at finite temperatures. In real films positional order may be stabilized by some substrate anisotropy, pinning or dipolar interaction. The last one is responsible for the emergence of magnetic-domain patterns in  films magnetized out of plane.     
In this context, Eq.~\eqref{eq67} may be applied to study the temperature dependence of the optimal period of modulation for a stripe pattern, known to be the 
ground state~\cite{Whitehead_PRB_95,Pighin_JMMM_10,Giuliani_PRB_07,Giuliani_PRB_11} (gray and white domains in Fig.~\ref{fig5}). 
At zero temperature, a modulated phase results from the competition between magnetostatic energy (which favors in this case antiparallel alignment of spin pairs) and the energy cost to cerate DWs. 
Several approaches lead to the same scaling of the characteristic period of modulation~\cite{Whitehead_PRB_95,Pighin_JMMM_10,Oliver_PRB_10}: 
$L_{\rm stripes} \sim {\rm e}^{\mathcal E_{\rm dw} /(4\Omega\tau^2)}$, with 
$\Omega=\mu_0 M_{\rm s}^2\mathfrak{a}^3/4\pi$ being the strength of dipolar interaction, $\tau$ the adimensional film thickness, $\mathfrak{a}$ the \textit{dimensional} lattice constant and 
$M_{\rm s}$ the saturation magnetization.  (remember that in this Section it is $\mathcal E_{\rm dw}=2\tau\sqrt{2J_{\rm atom}D_{\rm atom}}$).  
A first-order estimate of thermal effects is obtained by replacing $\mathcal E_{\rm dw}$ in the expression of $L_{\rm stripes}$
with $\mathcal F_{\rm dw}$ defined in  Eq.~\eqref{eq67}: $L_{\rm stripes} \sim {\rm e}^{\mathcal F_{\rm dw} /(4\Omega\tau^2)}$. 
The latter is consistent with a decrease of the period of modulation with increasing temperature, as observed in experiments~\cite{Oliver_PRL_06,Niculin_PRL_10}
and predicted by mean-field approach~\cite{Ale_PRB_08,Oliver_PRB_10}.   
Note that in deducing the DW free energy $\mathcal F_{\rm dw}$ dipolar interaction was totally overlooked. For instance, this interaction is known to produce an effective intermediate anisotropy that stabilizes Bloch type of DWs in films magnetized out of plane. Strictly speaking, since we assumed $D_x=0$, our treatment should not apply to 
stripped patterns emerging in those films. A more accurate study of this specific phenomenon -- which we intend to address in a forthcoming paper -- shall possibly produce a different entropic contribution to the DW free energy.  
Nevertheless, the main message of this paragraph is expected to hold true: Gaussian fluctuations around a DW profile suffice to account for a decrease of $L_{\rm stripes}$ with increasing temperatures.  \\

Still referring to domain patterns in films with dominant out-of-plane anisotropy, elastic deviations from the ideal stripe phase are known to yield anisotropic decay of correlations~\cite{Ar_Abanov_PRB_95}  
along $x$ and $y$. 
In the naive description of the stripe pattern given above, we ideally froze the elastic modes associated with compression along $x$. Yet, some disorder   
is expected to arise from the corrugation modes, developing along $y$ and associated with $\mathcal{Z}_{x_0}$. 
Without entering the details, this fact already suggests the decay of spatial correlations to be more severe along $y$ rather than along $x$. 
Consistently with this picture, several theoretical works~\cite{Ar_Abanov_PRB_95,Kashuba_PRB_93,Barci_PRE_13,Sergio_PRB_06} 
predict that the stripped ground state should evolve into a 2d smectic or nematic phase~\cite{Nelson_PRB_81} at finite temperature.  \\
The degeneracy with respect to $x_0$ of a single DW embedded in a thin film propagates to more complex patterns and leads to the \textit{floating-solid} description 
(the specific choice of $\vec n(x)$ being dependent on $x$ \textit{only} does not allow for rotational invariance on the $xy$ plane, which in physical systems also occurs).  \\
Usually, in real films all these effects are hindered by pinning. 
As related to the latter, the assumption -- stated by Hamiltonian~\eqref{eq55} -- that DWs behave as elastic interfaces is the staring point for describing depinning and creep dynamics~\cite{Brazovskii-Nattermann_AdvPhys_04,Giamarchi_PRL_98,Kim-Lee_Nature_09}.   
However, at relatively high temperatures, pinning may become negligible, thus restoring the idealized theoretical picture
sketched above. The observation of stripe mobility in Fe/Cu(001) films, indeed, supports this scenario~\cite{Oliver_PRL_06}. 
\subsection{Chiral order of domain walls} 
The instability with respect to $\varphi_0$ is, instead, related to DW chirality. 
As already pointed out, $\mathcal{Z}_{\varphi_0}$ may be thought of as arising from the 1d XY Hamiltonian given in Eq.~\eqref{eq61} in the limit of small misalignment between neighboring 
chirality vectors $\vec e_{\varphi_0}(y)$.   
It is well-known that the 1d XY model -- with nearest-neighbor interactions --  can only sustain short-range order. Therefore, our picture suggests that just short-range chiral order should develop along the $y$ direction when only uniaxial anisotropy and exchange interaction are considered. An intermediate anisotropy $D_x\ne0$ -- for instance of magnetostatic origin -- stabilizes two possible values of $\varphi_0$. 
Eventually, this drives the original XY Hamiltonian $\mathcal{H}_{\rm xy}$ in Eq.~\eqref{eq61}, which describes the \textit{effective} coupling between chirality vectors, towards the Ising universality class. 
Neither in this case long-range chiral order along $y$ is expected to be stable. From a snapshot taken at finite temperature we would rather expect domains of opposite chirality 
to alternate \textit{randomly} along, e.g., a Bloch DW (for which $\vec e_{\varphi_0}=(\pm1,\,0,\,0)$).  
This phenomenon was recently observed on Fe/Ni/Cu(001) films~\cite{Wu_PRL_2013}, consisting of 10 monolayers of Ni and 1.3 of Fe. Our conjecture is sketched pictorially in 
Fig.~\ref{fig5}. Arrows represent the alternating magnetization direction along DWs, instead of the $\vec e_{\varphi_0}$ vectors, to facilitate the comparison with experiments
reported in Ref.~\onlinecite{Wu_PRL_2013}.  
A scenario consistent with chiral order of DWs requires either a significant film thickness (so that dimensional crossover may occur) or the presence of a Dzyaloshinskii-Moriya interaction~\cite{Wiesendanger_Nature_07}.  
The latter, still in Ref.~\onlinecite{Wu_PRL_2013}, was observed to stabilize homochiral N\'eel type of DWs, with $\vec e_{\varphi_0}=(0,\,1,\,0)$, in samples with thinner Ni interlayer.   

Our considerations about short-range chiral order of DWs seem in striking contrast with Villain's conjecture~\cite{Villain_conj_78}  
(confirmed by experiments on Gd-based spin chains~\cite{Cinti_PRL_08}). This prescribes that in spin chains that develop short-range chiral order and are packed in a 3d crystal long-range chiral order should set in at higher temperature than magnetic ordering. 
In our mindset, the ``information'' about chirality of DWs in a 2d stripe-domain pattern cannot propagate along $x$, from one DW to the next because they are separated by regions in which all spins are aligned along the easy axis. 
Some misalignment between neighboring spins is, instead, needed to have a finite chiral order parameter. 
This ceases to hold true, e.g., close to the spin-reorientation transition where a canted-stripe phase was recently observed in 2d simulations~\cite{Whitehead_PRB_08,Pighin_PRE_12}.    
  
One should not forget that the arguments provided here strictly rely on thermodynamic equilibrium. Homochirality of DWs may be observed in experiments~\cite{Niculin_PRB_10,Venus_PRB_10,Venus_PRB_11} and simulations as a result of slow dynamics~\cite{Ana_PRE_13,Cannas_PRB_03,Kivelson_Phys_A_95}, similarly to what happens in superparamagnetic nanoparticles~\cite{Brown63PR,Brown79IEEE,Cheng06PRL}, for which long-range ferromagnetic order would be forbidden by equilibrium thermodynamics. 

Chirality is also related to adiabatic spin transfer torque (STT), through which an electric current may displace a DW hosted in a ferromagnetic nanowire. 
Translation of the DW, i.e. a variation of the $x_0$ parameter, is in this case necessarily accompanied by a precession of the $\varphi_0$ angle, which produces a  periodic change of the DW structure between Bloch and N\'eel type.  
The corresponding Landau-Lifshitz-Gilbert equation reads
\begin{equation}
\label{eq69}
\frac{{\rm d} \vec S}{{\rm d} t} = \gamma_0 \vec H_{\rm eff}\times\vec S + \alpha_{\rm G} \vec S\times \frac{{\rm d} \vec S}{{\rm d} t} - u \,\partial_x \vec S  
\end{equation}
where the effective field is defined from the Hamiltonian $\mathcal H$ in Eq.~\eqref{eq46} (possibly modified to include the magnetostatic energy or an intermediate anisotropy) as 
\begin{equation}
\label{eq70}
\vec H_{\rm eff} = - \frac{1}{g\mu_{\rm B}} \frac{\delta\mathcal H}{\delta \vec S } \,,
\end{equation}
$\gamma_0$ is the gyromagnetic ratio, $\alpha_{\rm G}$ the Gilbert damping and the last term accounts for the adiabatic STT with 
\begin{equation}
\label{eq71}
u = \frac{g\mu_{\rm B} J_{\rm e} P}{2e M_{\rm s}\mathfrak{a}} \,;
\end{equation}
in the equation above $J_{\rm e}$ represents the electrical current density and $P$ the polarization factor of the current (the lattice unit $ \mathfrak{a}$ has been added because the derivative in Eq.~\eqref{eq69} is considered dimensionless). 
In fact, the adiabatic STT contribution to the Landau-Lifshitz-Gilbert equation may be derived directly form the functional derivative in Eq.~\eqref{eq70} if a contribution  
\begin{equation}
\label{eq72}
\mathcal H_{\rm STT} = -\int\limits_0^{L_y+1} \mathrm dy \int\limits_0^{L_x+1} \vec \Gamma_{\rm STT}\cdot\left(\vec S \times \partial_x\vec S\right) \mathrm dx  
\end{equation}
is added to the Hamiltonian in Eq.~\eqref{eq46}. By analogy with standard Zeeman energy, Hamiltonian~\eqref{eq72}  alone would describe a precession of the local chirality vector~\cite{B_Braun_AdvPhys_2012} $\vec S \times \partial_x\vec S$ about and effective ``field'' $\vec \Gamma_{\rm STT}=\tau u \vec S$. 
Due to its close relation with chirality, we believe that reconsidering adiabatic STT in the perspective of short-range chiral order may contribute to shed some light on the puzzling scenario of DW motion induced by electric currents. For instance, adiabatic STT seems to catch the main physics of Co/Ni nanowires magnetized out of plane~\cite{Koyama_Nat_Mat_11,Koyama_IEEE_11}, 
while it largely fails for prototypical Permalloy nanowires (magnetized in plane)~\cite{Thiaville_JAP_04,Thiaville_EPL_05,Yang_PRB_08}; 
the critical current for DW motion is observed to depend strongly on temperature~\cite{Malinowski_JPD_10,Curiale_PRL_12} 
in some samples and not in others~\cite{Tanigawa_AppPhys_Ex_11}. 
Even if the extension to 2d of the sketch in Fig.~\ref{fig1} naturally leads to films or nanowires magnetized out-of-plane, our calculation can be adapted to samples magnetized in plane by a permutation of coordinates~\cite{A_Mougin_EPL_07,Boulle_Mat_SEng_11}. 
Within our picture, the ratio between the correlation length characterizing short-range chiral order 
($\xi_{\varphi_0}$) and the actual transverse size of a sample ($L_y$) should discriminate two regimes: for $\xi_{\varphi_0} \ll L_y$ the response of a DW to adiabatic STT is expected to depend strongly on temperature; for $\xi_{\varphi_0} \gg L_y$ this dependence should be much less dramatic. 
Also in this context, it is worth remarking that thermally-assisted DW depinning~\cite{Burrowes_NatPhys_10,Martinez_JPCM_12} 
and possible non-homogeneous mechanisms of precession~\cite{Zinoni_PRL_11} have been neglected in our considerations.  
\section{Conclusions}
We considered the effect of thermalized linear excitations about a DW profile. 
Expressions for the free energy of a DW embedded in 1d or 2d lattices were derived as a function of temperature and the system size. 
This was achieved by rephrasing in the language of Polyakov renormalization~\cite{Polyakov,Politi_EPL_94,Billoni_11} some known results, obtained from linearization of the 
Landau-Lifshitz equation~\cite{Winter_61,Mikeska_JPC_83,Yan_11,Yan_12,Hertel-Kirschner_PRL04,Bayer_05}.   
Our approach is equivalent to the steepest-descent approximation of functional integrals~\cite{Fogedby84JPCSSP}. It has, in our opinion, the advantage of allowing for an easier generalization to 2d.  
Moreover, it provides a better insight on the role of fast- and slow-varying degrees of freedom, while keeping track of the non-homogeneity within the fluctuation field. 
For instance, it is straightforward to realize that fluctuations associated with bound states shall be localized at the DW center. 
This information might be relevant to the aim of accounting efficiently for thermal fluctuations in the Landau-Lifshitz-Gilbert equation~\cite{wang_11}, beyond the mean-field level 
(Landau-Lifshitz-Bloch equation~\cite{Nowak_PRB_09}). 
   
From the knowledge of the DW free energy, we provided a phenomenological expansion for the correlation length that may be used to fit the susceptibility of Single-Chain Magnets 
(slow-relaxing spin chains~\cite{Miyasaka_review,Coulon06Springer,Bogani_JMC_08,Billoni_11,Gatteschi_Vindigni_13}). 
The last ones are often realized by creating a preferential path for the exchange interaction between anisotropic magnetic units -- consisting of transition metals or rare earths --  
through an organic radical. Thus, to some extent, Eq.~\eqref{eq1} can be considered a reference Hamiltonian for Single-Chain Magnets in general~\cite{Billoni_11,Gatteschi_Vindigni_13}.   

In a previous work~\cite{Ale_PRB_08}, we explained the shrinking of magnetic domains, observed in films with out-of-plane anisotropy, within a mean-field approach and assuming a non-homogeneous spin profile.  
Retaining the last feature for the unperturbed profile, in this paper we showed that Gaussian fluctuations also lead to a qualitatively similar result.  

As a further implication, our model suggests that long-range chiral order cannot occur within DWs interposed between saturated domains. 
The robustness of this conjecture certainly deserves to be checked \textit{beyond} the Gaussian approximation. 
Yet, it seems consistent with recent experiments on Fe/Ni/Cu(001) films~\cite{Wu_PRL_2013}. 
This softening of chiral order may acquire some relevance also in view of DW manipulations by means of spin-transfer torque.

\begin{acknowledgments}
A. V. would like to thank Lapo Casetti for stimulating discussions and Ursin Sol\`er for the valuable contribution in developing the code for TM calculations.    
We acknowledge the financial support of ETH Zurich and the Swiss National Science Foundation.  T.C.T.M acknowledges the financial support of St John's College, Cambridge. 
\end{acknowledgments}

\appendix
\section{Lowering and raising operators\label{operators}}
In the next two Appendices we provide some details about the solution of the eigenvalue problem~\eqref{eq6} in the presence of $\uparrow\downarrow$ b.c.. 
In terms of the lowering and raising operators
\begin{equation}
\begin{cases}
&\hat{a}_m  = \partial_\eta + m \tanh \eta \\
&\hat{a}^\dagger_m=-\partial_\eta +  m \tanh \eta \,, 
\label{eq22app}
\end{cases}
\end{equation}
the generalized fluctuation Hamiltonian given in Eq.~\eqref{eq21} reads $\hat{H}_m=\hat{a}^\dagger_m\hat{a}_m +m$. Using the commutator relation
\begin{equation}
[\hat{a}^\dagger_m,\hat{a}_m  ] = -\frac{2m}{\cosh^2\eta} \,,
\end{equation}
one finds the alternative representation $\hat{H}_m=\hat{a}_{m+1}\hat{a}^\dagger_{m+1} -(m+1)$.
We now show that if $|\psi\rangle_m$ is an eigenstate of $\hat{H}_m$ then
$\hat{a}^\dagger_{m+1} |\psi\rangle_m$ is an eigenstate of $\hat{H}_{m+1}$. Let $|\psi\rangle_m$ be an eigenstate of $\hat{a}_{m+1}\hat{a}^\dagger_{m+1} $ with eigenvalue $\lambda_m$, i.e. $\hat{a}_{m+1}\hat{a}^\dagger_{m+1} |\psi\rangle_m =\lambda_m |\psi\rangle_m$, then
\begin{align}
\hat{a}^\dagger_{m+1}(\hat{a}_{m+1}\hat{a}^\dagger_{m+1} |\psi\rangle_m)  & =\hat{a}^\dagger_{m+1}\hat{a}_{m+1}(\hat{a}^\dagger_{m+1} |\psi\rangle_m)\\
& =\lambda_m( \hat{a}^\dagger_{m+1}|\psi\rangle_m)\,,
\end{align}
which means that $\hat{a}^\dagger_{m+1}|\psi\rangle_m$ is an eigenstate of $\hat{a}^\dagger_{m+1}\hat{a}_{m+1}$ with the same eigenvalue $\lambda_m$. 
With the above representations of $\hat{H}_m$ and $\hat{H}_{m+1}$, it can easily be checked that if $|\psi\rangle_m$ is an eigenstate of $\hat{H}_m$ with eigenvalue $\mathcal{E}_m$ then $\hat{a}^\dagger_{m+1} |\psi\rangle_m$ is an eigenstate of $\hat{H}_{m+1}$ with eigenvalue $\mathcal{E}_{m}+2(m+1)$. Thus, the Hamiltonians $\hat{H}_m$ and $\hat{H}_{m+1}$ share the same spectrum, except for the fact that $\hat{H}_{m+1}$ has the additional eigenstate $|0\rangle_{m+1}$, defined by $\hat{a}_{m+1}|0\rangle_{m+1}=0$, which is not present in the spectrum of $\hat{H}_{m}$. This property allows constructing eigenstates of the Hamilton operator $\hat{H}_{m}$  recursively.
For the case of our interest, we stop the iteration at the first step $m=1$, namely we just apply $\hat{a}^\dagger_{1}$ to the free-particle eigenstates (see the following Appendix for the explicit calculation).  
The missing \textit{vacuum} state $|0\rangle_{1}$ is nothing but the bound state represented through Eq.~\eqref{eq24} in real space. 
\section{Derivation of the density of states\label{rho}}
\begin{figure}[hptb]
\centering
\input{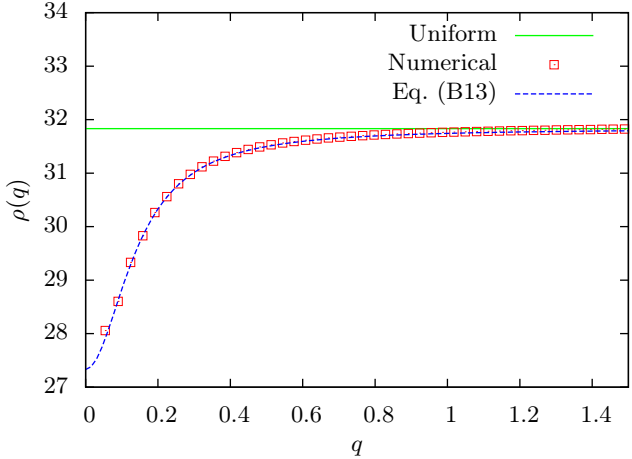} 
\caption{(Color online). Density of states $\rho(q)=\partial n/\partial q$ for $D/J=0.01$ and $L=99$. 
For small values of $q$ the density is decreased with respect to $\rho_{\uparrow\uparrow}=100/\pi$, expected for a 
uniform $\vec n$, due to DW--spin wave interaction. 
\label{fig:discrete_num}}
\end{figure}

In the finite system, the solutions of
the free--particle Hamiltonian are $\psi_0(q,x)=A\cos(qx)+B\sin(qx)$,
where $A\,,B$ are constants to be determined from the boundary and
normalization conditions. The \textit{free--state} solutions of the Schr\"odinger--like 
Eq.~\eqref{eq6} obtained for antiperiodicity b.c. (namely when $\vec n(x)$ describes a DW profile) can be obtained by applying 
the raising operator  $\hat{a}^\dagger_1$ defined in Eq.~\eqref{eq22} to $\psi_0(q,x)$:
\begin{equation}
  \label{eq:337}
  \begin{split}
  \Phi_{a,q}(x)=&A\left[q\sin(qx)/c+\tanh(c(x-x_0))\cos(qx)\right]\\
    &+B\left[\tanh(c(x-x_0))\sin(qx)-q\cos(qx)/c\right]\,, 
  \end{split}
\end{equation}
with $a=x_0,\varphi_0$. 
Independently of the determination of $A$ and $B$, it is straightforward to show that $\Phi_{a,q}(x)$ are solutions of the Schr\"odinger-like Eq.~\eqref{eq6} with eigenvalues 
\begin{equation}
  \label{eq16App}
  \varepsilon_{a,q}=\frac{J}{2}q^2+ D\,,
\end{equation}
formally equal to those obtained starting from a uniform profile ($\uparrow\uparrow$ b.c.). We show in the following that the allowed values of $q$ are not the same, 
which  affects the density of states. \\ 
The constants  $A$ and $B$ in Eq. \eqref{eq:337} have to be determined from the boundary conditions  $\Phi_{a,q}(0)= \Phi_{a,q}(L+1)=0$  and the normalization condition
\begin{equation}
  \label{eq:339}
  \int\limits_{0}^{L+1}\mathrm dx\,| \Phi_{a,q}(x)|^2=1\,.
\end{equation}
Though deriving an analytic expression for $A$ and $B$ is computationally demanding, the density of states can easily be obtained. Non-trivial solutions ($A\ne0$ and $B\ne0$) exist only if the 
determinant of the system defined by the boundary conditions $\Phi_{a,q}(0)= \Phi_{a,q}(L+1)=0$ vanishes: 
\begin{equation}
  \label{eq:340}
  \begin{split}
&\tanh(cx_0)\tanh\left(c(L+1-x_0) \right)\tan{q(L+1)}\\
&-q\tanh(cx_0)/c -q^2\tan{c(L+1-x_0)}/c^2\\
&-q\tanh\left(c(L+1-x_0) \right)/c=0\,,
  \end{split}
\end{equation}
which gives the following transcendental equation for the
determination of the possible $q$ values
\begin{equation}
  \label{eq:354}
  \tan{q(L+1)}=\frac{q\left(\tanh(cx_0)+\tanh\left(c(L+1-x_0)
\right)\right)}{c\left(\tanh(cx_0)\tanh\left(c(L+1-x_0)
\right)-q^2/c^2\right)}\,.
\end{equation}
This equation can be solved using, e.g., graphical methods. More significantly, it allows deriving an 
analytic expression for the density of states. To this purpose, let us set
\begin{equation}
  \label{eq:355}
  \tan{\alpha}=\frac{q\left(\tanh(cx_0)+\tanh\left(c(L+1-x_0)
\right)\right)}{c\left(\tanh(cx_0)\tanh\left(c(L+1-x_0)\right)-q^2/c^2\right)}\,.
\end{equation}
The solution of Eq. \eqref{eq:354} is
\begin{equation}
  \label{eq:356}
  \begin{cases}
  &  q(L+1)=\alpha+p\pi   \qquad\mathrm{with} \quad p\in\mathbb Z \\
  &  \alpha=\arctan\left[\frac{q\left(\tanh(cx_0)+\tanh\left(c(L+1-x_0)\right)\right)}
		{c\left(\tanh(cx_0)\tanh\left(c(L+1-x_0)\right)-q^2/c^2\right)}\right] 
  \end{cases}
\end{equation}
which can be solved for $p$ to obtain
\begin{equation}
\begin{split}
  \label{eq:357}
&  p=\frac{q(L+1)}{\pi} \\
&-\frac{1}{\pi}\arctan\left[\frac{q\left(\tanh(cx_0)
        +\tanh\left(c(L+1-x_0)
\right)\right)}{c\left(\tanh(cx_0)\tanh\left(c(L+1-x_0)
\right)
        -q^2/c^2\right)}\right]\,.
\end{split}
\end{equation}
The density of states~\cite{Currie_PRB_80} is defined as 
{\small
\begin{equation}
  \label{eq:362}
  \begin{split}
  \rho(q)&=\frac{{\rm d} p}{{\rm d}  q}=\frac{L+1}{\pi}-\frac{c}{\pi}\left[\tanh(cx_0)\tanh\left(c(L+1-x_0)
  \right)+  \frac{q^2}{c^2}\right]\\
  &\times\frac{\tanh(cx_0) + \tanh\left(c(L+1-x_0) \right)}{\Delta(q,x_0)}\,,  \qquad\mathrm{with}\\
  &\Delta(q,x_0) = c^2\left(\tanh(cx_0)\tanh\left(c(L+1-x_0)\right)-\frac{q^2}{c^2}\right)^2 \\
  &+ q^2\left[\tanh(cx_0)+ \tanh\left(c(L+1-x_0) \right)\right]^2\,.
  \end{split}
\end{equation}
}
In order to compute the DW free energy analytically, we assume the DW to be centered in the system, i.e., $x_0=(L+1)/2$. 
In this case, Eq. \eqref{eq:354} reads
  \begin{equation}
    \label{eq:363}
    \tan[q(L+1)]=\frac{2q\tanh\left(c(L+1)/2\right)}{c\left(\tanh^2(c(L+1)/2)-q^2/c^2\right)}\,.
  \end{equation}
  Setting $\tan\alpha=q/c\tanh(c(L+1)/2)$, Eq. \eqref{eq:363} takes the form
  \begin{equation}
    \label{eq:364}
    \tan[q(L+1)]=\tan(2\alpha)\,.
  \end{equation}
The density of states simplifies to 
\begin{equation}
  \label{eq:359}
 \rho(q)=\frac{L+1}{\pi} -\frac{2c\tanh(c(L+1)/2)}{\pi(c^2\tanh^2(c(L+1)/2)+q^2)}\,,
\end{equation}
and for $L\gg1/c$, $\tanh^2(c(L+1)/2)\approx1$,
\begin{equation}
  \label{eq:360}
  \rho(q)=\frac{L+1}{\pi}-\frac{2c}{\pi(c^2+q^2)}\,. 
\end{equation}
In order to check the validity of Eq.~\eqref{eq:359}, we performed a numerical diagonalization of  Eq.~\eqref{eq6} for a finite, discrete lattice 
(the excellent agreement is summarized in Fig.~\ref{fig:discrete_num}). \\

\textit{Diagonalization of the discrete eigenvalue problem \label{sec:discrete-case}} \\
The discrete version of Eq.~\eqref{eq6} is given by  
\begin{equation}
  \label{eq:348}
  \sum_{j,k}\psi(j)\left[-\frac{J}{2}\Delta_{j,k}+V(k)\delta_{j,k}\right]\psi(k)=\varepsilon\,, 
\end{equation}
where $\Delta_{j,k}$ is the discrete Laplace operator 
\begin{equation}
  \label{eq:163}
  \Delta_{j,k}=\begin{cases}
    -2 & j=k\\
    1 & k=j\pm1\\
    0 & \mathrm{else}
  \end{cases}\,,
\end{equation}
the potential is 
\begin{equation}
 \label{eq:163b}
  \begin{cases}
  V(k)=D\left(2\cos^2(\theta_k)-1\right) &\quad {\rm for }\quad \uparrow\downarrow  \,{\rm b.c.} \\
  V(k)=D& \quad {\rm for }\quad  \uparrow\uparrow\,{\rm b.c.}\,,
  \end{cases}
\end{equation}
and $\theta_k$ corresponds to the discrete profile computed, e.g., with the 
non-linear map method (see Appendix~\ref{finite_size_DW}). 
A complete orthonormal system fulfilling the boundary
conditions $\psi(0)=\psi(L+1)=0$ is
\begin{equation}
  \label{eq:349}
  u_n(j)=\sqrt{\frac{2}{L+1}}\sin(q_n j)\qquad\qquad q_n=\frac{\pi}{L+1}n
\end{equation}
with $0\leq j\leq L+1$ and $1\leq n\leq L$. 
The functions $\psi$ are expanded in this basis
\begin{equation}
  \label{eq:350}
  \psi(j)=\sum_n w_n u_n(j)
\end{equation}
and inserted into Eq. \eqref{eq:348} to get
\begin{equation}
  \label{eq:351}
  \sum_{n,m}w_n\left[J(1-\cos(q_n))\delta_{n,m}+M_{n,m}\right]w_m=\varepsilon\,,
\end{equation}
with
\begin{equation}
  \label{eq:353}
  M_{n,m}=\sum_ju_n(j)V(j)u_m(j)\,.
\end{equation}
This is equivalent to solving the eigenvalue problem
\begin{equation}
  \label{eq:352}
  \sum_m\left[J(1-\cos(q_m))\delta_{n,m}+ M_{n,m}\right]w_m=\varepsilon_n w_n\,.
\end{equation}
It is worth noting that in a finite discrete system  the possible number of
eigenfunctions $L$ is finite. As a consequence, every time a domain wall is
added, a free state is ``lost'' and a bound state is
``gained''. However, this procedure is correct only when the distance
between DWs is large enough to treat them
independently. In our case we just considered one or no DW in the system. 
When no DW is present, i.e. for $\uparrow\uparrow$ b.c., $u_n(j)$ are solutions  
to the eigenvalue problem in Eq.~\eqref{eq:348} with eigenvalues
\begin{equation}
  \label{eq:171}
  \varepsilon_{\uparrow\uparrow,n}= J(1-\cos(q_n)) +D \,.
\end{equation}
As already mentioned, the dispersion relation is expected to depend on $q$ in the same way for both choices of b.c., but the allowed values of $q$ may -- generally -- be different.  
The spectrum in Eq.~\eqref{eq:171} was compared with that obtained by solving the eigenvalue problem~\eqref{eq:352} numerically:  
$\varepsilon_n$ obtained in both cases were plotted against the eigenvalue index $n$. Indeed, the two dispersion relations turned out to overlap. 
Therefore, the ``shifted'' $q_{\uparrow\downarrow,n}$, corresponding to $\uparrow\downarrow$ b.c.,   could be deduced through the following formula:
\begin{equation}
  \label{eq:341}
  q_{\uparrow\downarrow,n}=\arccos\left(\frac{J+ D-\varepsilon_{\uparrow\downarrow,n}}{J}\right)\,.
\end{equation}
Eventually, the density of states was obtained with the discrete derivative
\begin{equation}
  \label{eq:342}
\rho(q)=\frac{\partial n}{\partial q} \simeq\frac{1}{q_{\uparrow\downarrow,n+1}-q_{\uparrow\downarrow,n}}\,.
\end{equation}
Symbols in Fig.~\ref{fig:discrete_num} represent the resulting density of states.  
Analytic results are drawn with lines: the (blue) short-dashed line corresponds to Eq.~\eqref{eq:360} while the  (green) solid line to $\rho_{\uparrow\uparrow}=(L+1)/\pi$. 
Indeed,  Eq.~\eqref{eq:360}  agrees very well with the numerical calculation. 
\section{Scaling of the correlation length \label{xi}}
In this Appendix we describe the way in which a semi-analytical expression for the correlation length was deduced.  
As mentioned in the main text and shown in Fig.~\ref{fig3}, a better agreement between $\Delta F$ given in Eq.~\eqref{eq31} and numerical results is obtained 
by introducing the renormalized parameters $D(\lambda_{\rm c})$ and 
$c(\lambda_{\rm c})$ \textit{only} in the argument of the logarithm. Considering that $D(\lambda_{\rm c})=D\, Z^3(\lambda_{\rm c})$ 
and $c(\lambda_{\rm c})=c\, Z(\lambda_{\rm c})$, after this substitution, the DW free energy takes the form 
\begin{equation}
\label{App_xi_1}
\Delta F= \mathcal E_{\rm dw} - T\log\left\{\beta \mathcal E_{\rm dw} c L  Z^3 \left[1+\tanh(cZ(L+1)/2)\right]^2\right\}\,. 
\end{equation}
We are interested in the value of $L=\bar L$ for which $\Delta F(\bar L)=0$. Let us set  
$\tilde{\xi}=c\bar L$ and $\zeta=\beta \mathcal E_{\rm dw}$ so that our problem reduces to an implicit equation in the 
variables $\tilde{\xi}$ and $\zeta$. Then consider the following map in the $\tilde{\xi}$ variable 
\begin{equation}
\label{App_xi_2}
\begin{split}
&\Xi_{it}=\frac{Z(\zeta)}{2}(\tilde{\xi}_{it}+c)\\
&\tilde{\xi}_{it+1}=\frac{{\rm e}^\zeta}{\zeta Z^3(\zeta)}\left[1+\tanh(\Xi_{it})\right]^{-2}
\end{split}
\end{equation} 
parametrically dependent on $\zeta$, with $it=1,2,\dots$ representing the iteration index. From its definition it follows that $Z(\lambda_{\rm c})=Z(\zeta)=1-1/\zeta$. 
The initial condition of the map~\eqref{App_xi_2} is set to
\begin{equation}
\label{App_xi_3} 
\tilde{\xi}_{1}=\frac{{\rm e}^\zeta}{4\zeta Z^3(\zeta)}\,.
\end{equation}
The fixed point $\tilde{\xi}_{\infty}$ gives the sought for solution  $\Delta F(\bar L)=0$. 
For $\tilde{\xi}_{1}\gg1$, it is $\Xi_{1}\gg 1$ and $\tanh(\Xi_{1})\simeq 1$, consequently. In this case the map converges already at the first iteration, namely 
$\tilde{\xi}_{1}\simeq \tilde{\xi}_{\infty}$. Numerically one finds that the latter equivalence holds within 1.4\% for any $D/J>0.1$. \\
To extend the validity of Eq.~\eqref{App_xi_3} to high temperatures we used the trick -- not fully justified -- that Polyakov renormalization 
somehow stops when $\lambda_{\rm c}=\bar L/2 - 3/2c$. In terms of the reduced variable entering the map~\eqref{App_xi_2}, this happens when 
$\tilde{\xi}_{\infty}=2\pi+3$, which occurs at $\zeta=4.262$.  We are now in the position to 
give an expression for the scaling function $f(\zeta)$ describing the \textit{universal} behavior of the product $c\,\xi$ as a function of the scaling variable 
$\zeta =\beta \mathcal E_{\rm dw}$:
\begin{equation}
\label{App_xi_4}
f(\zeta)=
\begin{cases}
\frac{{\rm e}^\zeta}{4\zeta Z^3(\zeta)}		&\quad {\rm for } \,\,\, \zeta>4.262\\
& \\
A_0 \, {\rm e}^{b \zeta + 1} 	&\quad {\rm otherwise} 
\end{cases}
\end{equation}  
with constants $A_0= 0.3282$  and  $b= 0.5496$ determined by requiring the continuity of $\tilde{\xi}_{1}$ and its derivative with respect to $\zeta$. 
The exact relation between the scaling function $f(\zeta)$ and $\xi$ computed numerically can only be obtained by fitting a constant prefactor (see Fig.~\ref{fig3}); we already pointed out that this is of the order of the Euler constant so that we can reasonably set the equivalence $c\,\xi\simeq f(\zeta)/{\rm e}$.  
\section{Spin profile in laterally constrained domain walls\label{finite_size_DW}}
Here we repropose an analytic derivation of the DW profile for finite systems~\cite{B_Braun_AdvPhys_2012,B_Braun_JAP_2006} -- which employs a continuum formalism -- 
and check its results against numerical calculations on a discrete lattice. \\
We first consider the DW profile for an infinite chain with $\uparrow\downarrow$ b.c.. 
Since for $T=0$ one has $\vec n(x) = \vec S(x)$, the minimum-energy profile given in Eq.~\eqref{eq17} can be obtained from Hamiltonian~\eqref{eq2}.  
The latter in polar coordinates 
$\vec S \equiv (\sin\theta\cos\varphi,\sin\theta\sin\varphi,\cos\theta)$ reads  
\begin{equation}
 \label{eq:218}
 \begin{split}
   \mathcal H &=\int \frac{J}{2}\left[\left(\partial_x \theta \right)^2+\sin^2(\theta)\left(\partial_x\varphi \right)^2\right]\,\mathrm dx\\
&-\int \left[ D\cos^2(\theta)+D_x\sin^2(\theta)\cos^2(\varphi) \right]\,\mathrm dx  +\mathrm{const} \,.
\end{split}
\end{equation}
The intermediate anisotropy $D_x$ such that $0<|D_x|<|D|$ has been introduced for the sake of generality.  
The profile which minimizes the functional in Eq.~\eqref{eq:218} with respect to $\theta(x)$ and $\varphi(x)$ 
is the solution to the following Euler--Lagrange equations
\begin{equation}
 \begin{split}
  \label{eq:219}
&\bullet\quad  J\partial_x^2\theta=J\sin\theta\cos\theta\left(\partial_x\varphi\right)^2+ 2\tilde D\cos\theta\sin\theta\\
&\bullet\quad  J\sin^2(\theta)\partial_x^2\varphi + 2J\sin\theta\cos\theta\partial_x\theta\partial_x\phi \\
&\quad\quad=2D_x\sin^2(\theta)\sin\varphi\cos\varphi
\end{split}
\end{equation}
(where $\tilde D=D-D_x\cos^2(\varphi)$) compatible with $\uparrow\downarrow$ b.c., 
that is 
\begin{equation}
\label{eq:222}
\begin{split}
&\cos\left(\theta(x)\right)=-\tanh(c(x-x_0))\\
&\varphi(x)=\varphi_0=0,\frac{\pi}{2},\pi,\frac{3\pi}{2}\,,
\end{split}
\end{equation}
with the more general $c=\sqrt{2\tilde D/J}$ than in the main text. The choice of $\varphi_0$ depends on the sign of the intermediate anisotropy. For $D_x>0$ one has
$\varphi_0=0,\pi$ (N\'eel DW), while for $D_x<0$ it is $\varphi_0=\pi/2,3\pi/2$ (Bloch DW). 
Note that in both cases there exist two degenerate solutions which correspond to opposite chirality of the DW~\cite{B_Braun_AdvPhys_2012}.  
The energy associated with the profile \eqref{eq:222} -- with respect to a uniform ground state -- is $\mathcal E_{\rm dw} =2\sqrt{2\tilde DJ}-D_x$.

\begin{figure}[hptb]
\centering
\input{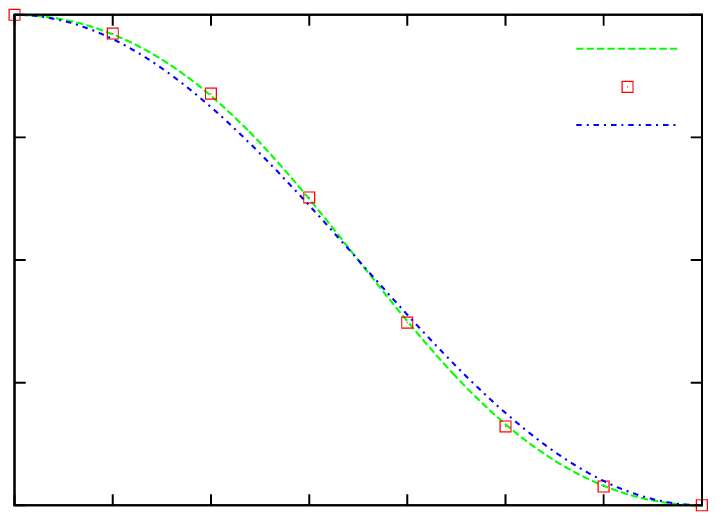}
\hfill
\input{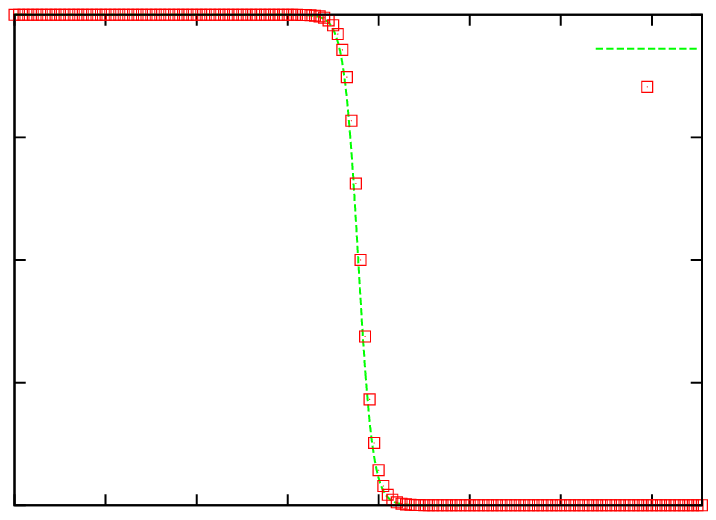}
\caption{(Color online). DW spin profile obtained with different methods for $D/J=0.05$, yielding $c^{-1}\approx3.16$. 
Non--linear map and continuum approach produce the same spin profile for both $L=6,\,150$.  
For $L=6$, the harmonic approximation given by Eq.~\eqref{eq:280} is also accurate.
\label{fig_profile}}
\end{figure}

A helpful, well-established~\cite{Leung_82} analogy consists in interpreting $x$ as ``time'' and  $\theta$ as ``spatial coordinate''. 
Then the  first Euler--Lagrange equation in~\eqref{eq:219} describes the motion of a classical particle of ``mass'' $J$ moving in a potential 
$V(\theta)= D\cos^2(\theta)+D_x\sin^2(\theta)\cos^2(\varphi_0)$ ($\varphi=\varphi_0$ has been assumed). 
The ``particle energy''    
\begin{equation}
 \label{eq:273}
 \varepsilon=\frac{J}{2}(\partial_x\theta)^2+\frac{\tilde D}{2}\cos(2\theta) 
\end{equation}
is a constant of integration and it is univocally defined by the b.c. 
\begin{equation}
  \label{eq:221}
  \begin{cases}
    &\theta(0)=0\\
    &\theta(L+1)=\pi\,,
  \end{cases}
\end{equation}
consistent with Fig.~\ref{fig1}. For the infinite chain $\theta(-\infty)=0$ and $\theta(\infty)=\pi$, which yields
$\varepsilon=D/2$, namely $\partial_x\theta=0$ at the boundaries. 
For finite chains, instead, one expects $\varepsilon>D/2$, i.e., a non-vanishing derivative at the boundaries. 
We will see that this condition is necessary to develop a continuum model for finite chains (it guarantees the convergence of the 
elliptic integrals in  Eqs. \eqref{eq:274}--\eqref{eq:278}). 
Integration of Eq.~\eqref{eq:273} gives the following implicit equation for the DW profile 
\begin{equation}
  \label{eq:274}
  x=\sqrt{\frac{J}{\tilde D}}\int\limits_0^\theta\,\frac{\mathrm
    d\theta'}{\sqrt{\frac{2\varepsilon}{\tilde D}-\cos(2\theta')}}
\end{equation}
where the b.c. $\theta(0)=0$ has been used. The second b.c. in Eq.~\eqref{eq:221} implies 
\begin{equation}
  \label{eq:278}
  L+1=2\sqrt{\frac{J}{2\varepsilon+\tilde D}}\, \, \mathcal K\left(\sqrt{\frac{2\tilde D}{2\varepsilon+\tilde D}}\right)\,,
\end{equation}
$\mathcal K$ being the complete elliptic integral of the first kind. 
This equation can be solved numerically to deduce $\varepsilon$. The whole model holds for broad DWs, $c \ll 1$, 
when the continuum limit is appropriate.  Numerical instabilities are expected 
for $\varepsilon\simeq\tilde D/2$, when integrals diverge. Assuming that the DW is centered at $(L+1)/2$ with $\theta((L+1)/2)=\pi/2$, the spin profile can also be computed as 
\begin{align}
\label{eq:279}
x=\frac{L+1}{2} - \sqrt{\frac{J}{\tilde D}}\int\limits_{\theta}^{\pi/2}\,\frac{\mathrm
d\theta'}{\sqrt{\frac{2\varepsilon}{\tilde D}-\cos(2\theta')}}
\end{align}
 which solves the possible numerical divergence at the 
boundaries. Though Eq.~\eqref{eq:279} only applies to $x\le(L+1)/2$, the other half of the profile can be deduced from the property 
$\theta(L+1-x)=\pi-\theta(x)$.  
For very short chains the constant of integration is approximately 
$\varepsilon=J(\partial_x\theta)^2/2$ 
(meaning that in Eq.~\eqref{eq:273} the derivative dominates over the cosine term also at boundaries). 
As a consequence, the condition \eqref{eq:278} takes the form
\begin{equation}
  \label{eq:380}
  L+1=\sqrt{\frac{J}{2\varepsilon}}\int\limits_0^\pi\mathrm d\theta\,,
\end{equation}
which gives $\varepsilon=J(\pi/(L+1))^2/2$, i.e., the
derivative $\partial_x\theta$ is constant over the whole chain and
takes the value of the smallest wave-vector $q$ allowed in the reciprocal space. The spin profile can be computed analytically
\begin{equation}
  \label{eq:280}
  \theta(x)=\frac{\pi}{L+1}x\quad\Rightarrow\quad S^z(x)=\cos\left(\frac{\pi}{L+1}x\right)\,.
\end{equation}
The above profile corresponds to a single harmonic. In Figure~\ref{fig_profile}, the spin profile obtained with different methods (at $T=0$) 
for two different chain lengths is shown. The harmonic approximation is satisfactory for $L=6$ but -- as expected -- does not reproduce the discrete-lattice calculation for $L=150$ (not shown). 
 \\

\textit{Discrete-lattice calculation } \\ 
In the case of a discrete chain, a recursion formula (non--linear map) for the
computation of the spin profile was proposed in Refs.~\onlinecite{trallori:94,rettori:95,trallori:96}. The starting point is the DW Hamiltonian
\begin{equation}
  \label{eq:281}
  \mathcal
  H_{\rm dw}=J\sum_{i=0}^L\left(1-\cos\left(\theta_{i+1}-\theta_i\right)\right)+
  \sum_{i=0}^{L+1}D\sin^2(\theta_i)\,,
\end{equation}
where the decomposition
\begin{equation}
  \label{eq:282}
  \vec S_i\cdot\vec S_j=\cos(\theta_i)\cos(\theta_j)+\sin(\theta_i)\sin(\theta_j)\cos(\varphi_i-\varphi_j)
\end{equation}
with $\varphi_i=\varphi_0$ was used. To find the profile, $\mathcal H_{\rm dw}$
has to be minimized
\begin{equation}
  \label{eq:283}
  \frac{\partial \mathcal
    H_{\rm dw}}{\partial\theta_i}=0\,,\qquad\qquad 0<i<L+1
\end{equation}
which yields
\begin{equation}
  \label{eq:284}
  J\sin(\theta_i-\theta_{i-1})- J\sin(\theta_{i+1}-\theta_i)+ D\sin(2\theta_i)=0\,.
\end{equation}
The non-linear map is built as follows
\begin{equation}
  \label{eq:285}
  \tilde s_{i+1}=\sin(\theta_{i+1}-\theta_i)\,,
\end{equation}
so that
\begin{equation}
  \label{eq:286}
  \tilde s_{i+1}=\tilde s_{i}+\frac{D}{J}\sin(2\theta_i)\,,
\end{equation}
and
\begin{equation}
  \label{eq:287}
  \theta_{i+1}=\theta_{i}+\arcsin(\tilde s_{i+1})\,.
\end{equation}
The procedure gives a unique solution, when the boundary conditions are inserted
\begin{equation}
  \label{eq:288}
  \theta_0=0\qquad\qquad\qquad\theta_{L+1}=\pi\,,
\end{equation}
and the first step is given, $\theta_L=\pi-\delta$. The difficulty of
the procedure is to find the correct value of $\delta$ which
satisfies the boundary conditions \eqref{eq:288}. In practice, one
tries different values of $\delta$ until $|\theta_{0}|$ is smaller than a preset threshold. However, this procedure gets 
numerically unstable for long chains. For $L$ odd, one can circumvent the problem by letting the iteration start at the 
DW center, where $\theta_{(L+1)/2}=\pi/2$ and $\theta_{(L+3)/2}=\pi/2+\delta$. The more dramatic misalignment between two adjacent spins occurs 
at the DW center. Therefore here the iteration step $\delta$ can be defined more easily.  
After a certain iteration $\bar i$, all the spins characterized by a lattice index $i>\bar i$
are assumed to point along the same direction as the boundary, $\theta_i=\pi$.  
The configuration of spins lying on the other side of the DW one has is obtained from the condition 
\begin{equation}
  \label{eq:290}
  \theta_{L+1-i}=\pi-\theta_i\qquad (L+1)/2<i\leq L+1
\end{equation}
Figure \ref{fig_profile} shows that the results obtained with the  non-linear map (for a discrete lattice) 
and those computed with the continuum formalism (by means of Eqs.~\eqref{eq:278} and~\eqref{eq:279}) are not distinguishable. 

\bibliographystyle{apsrev4-1}
\bibliography{Sangiorgio}
\end{document}